\documentclass[useAMS, usenatbib]{mn2e}

\usepackage{mathptmx}
\usepackage{graphicx}
\def\lsun{\rm L_{\odot}}
\def\msun{\rm M_{\odot}}

\def\sfr{\rm M_{\odot}\ yr^{-1}}

\def\micron{$\mu$m}
\title[Pre-processing in the local Universe]{From Voids to Coma: the prevalence of pre-processing in the local Universe}
\author[R.~Cybulski et al.]{Ryan Cybulski,$^{1,2}$\thanks{E-mail:
jcybulsk@astro.umass.edu} Min S.~Yun,$^{1}$ Giovanni G.~Fazio,$^{2}$ 
Robert A.~Gutermuth$^{1}$ \\
$^{1}$Department of Astronomy, University of Massachusetts, Amherst, MA 01003, USA\\
$^{2}$Harvard-Smithsonian Center for Astrophysics, 60 Garden Street, Cambridge, MA 02138, USA}
\begin{document}

\date{Accepted 2014 January 28. Received 2014 January 16; in original form 2013 November 19}

\pagerange{\pageref{firstpage}--\pageref{lastpage}} \pubyear{2014}

\maketitle

\label{firstpage}

\begin{abstract}
We examine the effects of pre-processing across the Coma Supercluster, including 3505 galaxies over $\sim$500 deg$^{2}$, by quantifying the degree to which star-forming (SF) activity is quenched as a function of environment. We characterise environment using the complementary techniques of Voronoi Tessellation, to measure the density field, and the Minimal Spanning Tree, to define continuous structures, and so we measure SF activity as a function of local density \textit{and} the type of environment (cluster, group, filament, and void), and quantify the degree to which environment contributes to quenching of SF activity. Our sample covers over two orders of magnitude in stellar mass (10$^{8.5}$ to 10$^{11}\msun$), and consequently we trace the effects of environment on SF activity for dwarf and massive galaxies, distinguishing so-called `mass quenching' from `environment quenching'. Environmentally-driven quenching of SF activity, measured relative to the void galaxies, occurs to progressively greater degrees in filaments, groups, and clusters, and this trend holds for dwarf and massive galaxies alike. A similar trend is found using $g-r$ colours, but with a more significant disparity between galaxy mass bins driven by increased internal dust extinction in massive galaxies. The SFR distributions of massive SF galaxies have no significant environmental dependence, but the distributions for dwarf SF galaxies are found to be statistically distinct in most environments. Pre-processing plays a significant role at low redshift, as environmentally-driven galaxy evolution affects nearly half of the galaxies in the group environment, and a significant fraction of the galaxies in the more diffuse filaments. Our study underscores the need for sensitivity to dwarf galaxies to separate mass-driven from environmentally-driven effects, and the use of unbiased tracers of SF activity.
\end{abstract}


\section{Introduction}

Studies of massive galaxy clusters and groups at $z\le1$ typically find environments with little-to-no star formation activity, in sharp contrast with the field. Over-dense regions are dominated by red, passively-evolving S0 and elliptical galaxies, whereas more sparsely-populated regions tend to have galaxies with spiral morphologies, younger stellar populations, and systematically higher star formation rates \citep*{dressler1980, postmangeller1984, pimbblet2002, poggianti2006, haines2007, gavazzi2010, mahajan2010, peng2010, scoville2013}. An observed trend of increasing blue galaxy fraction with redshift \citep[the `Butcher-Oemler' effect;][]{butcheroemler1984} has been interpreted as evidence for higher star formation activity and stellar mass build-up in higher redshift clusters -- or alternatively, that star formation is quenched more recently by one or more processes in over-dense regions.

Several physical mechanisms can account for the quenching of star formation in over-dense regions \citep[for a review, see][]{boselligavazzi2006}. Galaxies in environments with sufficiently low velocity dispersions can be strongly perturbed by mergers. Galaxies can also be transformed more gradually by an ensemble of small perturbations with neighbours, a process called harassment \citep{moore1999}. Tidal forces can strip away a galaxy's halo gas \citep[starvation;][]{larson1980, balogh2000, bekki2002}, cutting off a fuel source for future star formation and leading to a gradual decline in SF activity. In the high-density cores of massive clusters, the hot ($T_X\sim10keV$) intra-cluster medium (ICM) can quench star formation by removing gas from galaxies via ram-pressure stripping \citep{gunngott1972, abadi1999, quilis2000, kronberger2008, bekki2013}. The relative strengths of these physical mechanisms are strongly dependent on the cluster or group properties (dynamical state, mass, and intra-cluster or intra-group medium) and environment.

Targeted studies of galaxy clusters or groups at $z\leq1$ have revealed overwhelming evidence that galaxy transformation occurs not just in dense cluster cores, but at lower densities characteristic of cluster outskirts or galaxy groups \citep{zabludoffmulchaey1998}. Studies with star formation tracers in the IR \citep{fadda2000, marcillac2007, tran2009, bai2010, biviano2011}, UV \citep{just2011, lu2012, rasmussen2012}, and optical emission-line measures \citep{tran2005, poggianti2006} have shown evidence of \textit{pre-processing}, whereby infalling galaxies undergo changes prior to their arrival in the galaxy cluster, or galaxies are transformed entirely in the group environment \citep{fujita2004, bahe2013}. The pre-processing hypothesis has also been supported by studies of the environmental dependence on galaxy morphology \citep[][]{helsdonponman2003, poggianti2009b} and colour \citep{mok2013, trinh2013}. 

Numerical simulations have also been used to study the causes and implications of galaxy pre-processing. \citet{bekkicouch2011} showed that the dominant physical processes galaxies are likely subjected to in group environments, specifically the frequent weak tidal interactions of harassment, are capable of transforming late-type, disk-dominated galaxies into bulge-dominated, early-types. Furthermore, \citet{mcgee2009} used simulations of dark matter halo merger trees, with semi-analytic models (SAMs) to populate the haloes with galaxies, and traced the histories of the simulated galaxies that ended up accreting onto cluster-mass haloes in different epochs. In doing so, \citet{mcgee2009} determined what fraction of those cluster galaxies had resided in haloes characteristic of group-masses for a long enough time to have been pre-processed prior to entering the cluster. The results of their simulation showed that at low redshift a large fraction of cluster galaxies could have been affected by their environment prior to entering the cluster, while at earlier epochs the fraction of pre-processed galaxies in clusters should steadily decline. The fraction of cluster galaxies affected by pre-processing in the \citet{mcgee2009} simulation depends on the assumed timescale for the physical process(es) in group environments to affect galaxies, and also has a stellar mass dependence. Although many assumptions go into this simulation, the result highlights a key point that the role of pre-processing has likely varied significantly over cosmic time, and that at $z\sim0$ pre-processing should be extremely prevalent.

Recent studies have suggested that the quenching of SF activity in cosmic history is primarily driven by two distinct, and possibly separable, components: secular evolution (or `mass quenching') and environmentally-driven processes \citep[or `environment quenching';][]{baldry2006, peng2010}. However, see also \citet{delucia2012} for a discussion about how \textit{history bias} affects one's ability to disentangle mass- and environment-quenching. Nevertheless, any attempt to examine the environmental dependence on galaxy evolution must include a careful account for the possibility that one's galaxy selection function has mass biases, particularly since the galaxy stellar mass function is known to vary with environment \citep{cooper2010}. Concerns about biases introduced by the galaxy selection function are compounded when examining galaxies over a wide range in redshift, as one's sensitivity, in galaxy mass and in other properties, like SFR, will undoubtedly also vary with $z$. As a result, in many of these studies that extend to higher-$z$ one must restrict one's sample to only massive galaxies with high SFRs, and thereby have a less complete picture of the effects of environment on galaxy evolution. Furthermore, studies extending to higher-$z$ tend to sample a smaller dynamic range of environments, which similarly reduces one's ability to draw general conclusions about environmentally-driven processes. A comprehensive view of galaxy evolution in different environments must be sensitive to a large dynamic range of local densities in order to capture not just the dense regions, like clusters and groups, but the more diffuse filament and void regimes.

A key challenge faced when interpreting the many results examining galaxy evolution, in addition to the aforementioned sources of potential bias, is the wide range of methods employed to characterise environment. Recently, \citet{muldrew2012} used an array of different environmental mapping techniques, which could be roughly grouped into two categories: nearest-neighbour methods, which measure galaxy density with an aperture that changes depending on the local galaxy density, and fixed-aperture techniques, whose apertures do not vary, to examine a mock galaxy catalogue. \citet{muldrew2012} found that these techniques can analyse the same data set and get different results, but that the nearest-neighbour methods appear to be optimal for mapping the density fields within massive haloes, while the fixed-aperture methods are better suited for probing superhalo distance scales. Therefore, the technique that is optimal to identify large scale structures (LSS), like clusters, groups, and filaments, is not necessarily the best choice for measuring the density fields within those structures.

In this work, we seek to quantify the role of pre-processing in the local universe by analysing the rest-frame colour and star-formation activity of galaxies as a function of environment over about three orders of magnitude in projected density in the Coma Supercluster. By focusing on a low-$z$ field, we ensure that our sample of galaxies, taken from the Sloan Digital Sky Survey \citep[SDSS;][]{york2000}, is spectroscopically complete down to dwarf masses ($M_*\ge10^{8.5}\msun$). Furthermore, we do not have to rely on photometric redshift (photo-$z$) measurements, which would introduce additional contamination due to interlopers in our sample and significant smearing along the line-of-sight. To map the environments of the supercluster, we employ two complementary techniques: Voronoi Tessellation (VT) and the Minimal Spanning Tree (MST). The former is a nearest-neighbour-based approach, which can measure the local density field effectively over the large dynamic range of densities that we find in the Coma Supercluster. The latter technique is most effective at characterising continuous structures, like clusters, groups, and filaments, and therefore we use the MST to differentiate the types of environment extending over super-halo scales. Our combined VT and MST approach allows us to select discrete components of the cosmic web by exploiting the fundamental density contrasts of the cluster, group, filament, and void environments. Another benefit of the proximity of our target field is sensitivity to low SFRs, as our combined approach of using the {\it Galaxy Evolution Explorer} \citep[GALEX;][]{martin2005} and Wide-Field Infrared Survey Explorer \citep[WISE;][]{wright2010} to recover unobscured and dust-obscured star-formation activity, respectively, across the entire Coma Supercluster down to 0.02$\sfr$.

Section \ref{sample} describes the Coma Supercluster and our sample selection process, with our data from SDSS, {\it GALEX}, and WISE. In Section \ref{lss} we outline our techniques for mapping the LSS in the Coma Supercluster, and in Section \ref{results} we present our resulting SFRs and comparisons of SF activity and colour versus environment. In Section \ref{discussion} we discuss the implications of our results, and compare our work to previous studies. Throughout this paper we use cosmological parameters $\Omega_{\Lambda}=0.70$, $\Omega_{M}=0.30$, and $H_0=70$ km s$^{-1}$ Mpc$^{-1}$, where pertinent cosmological quantities have been calculated using the online Cosmology Calculator of E.~L.~Wright \citep{wright2006}. Throughout we assume a Kroupa IMF \citep{kroupa2001}, and hereafter we will refer to galaxies with stellar masses M$_*\leq$10$^{9.5}\msun$ as dwarf galaxies, and those with M$_*>$10$^{9.5}\msun$ as massive galaxies.

\section{Sample Selection}\label{sample}

The Coma Supercluster is an ideal field to observe signatures of galaxy transformation in different environments. It contains two rich galaxy clusters, Abell 1656 and Abell 1367, and several galaxy groups distributed in a filamentary pattern between the two clusters \citep{gregorythompson1978}. Furthermore, the two clusters are in very different dynamical states, with A1656 being relaxed and A1367 still undergoing significant merging \citep{donnelly1998, girardi1998, cortese2006}. The close proximity of the supercluster ($z\simeq0.023$) allows us to probe its galaxy population down to dwarf masses ($M_* \sim 10^{8.5} M_{\odot}$) with a spectroscopically complete sample, and the geometric alignment of the supercluster, with the galaxy distribution extending largely perpendicular to our line-of-sight \citep{chincarini1983}, makes it an ideal case study to examine galaxies in a wide range of environments with minimal projection effects.

Past studies of the Coma Supercluster have been primarily focused on the most massive cluster, A1656. Its low redshift, high galactic latitude ($b\sim88^\circ$), and richness ensured that it received a great deal of attention from observers in early extragalactic studies \citep[see][and references therein]{biviano1998}. A significant substructure $\sim$1 Mpc SW of the centre of A1656, which has since been positively identified as an infalling group \citep{neumann2001}, was noticed first by the high local concentration of galaxies centred on the galaxy NGC 4839, and was later confirmed by a diffuse X-ray profile and radial velocities of member galaxies. \citet{caldwell1993} found a large number of `post-starburst' (or k+A) galaxies coincident with the NGC 4839 group, leading to the conclusion that the NGC 4839 group had experienced a burst of star formation $\sim1$ Gyr ago \citep{caldwellrose1997}, possibly triggered by tidal effects of the group-cluster merging \citep{bekki1999}. A study by \citet{poggianti2004}, examining emission-line and k+A galaxies in the core region of A1656, found a spatial correlation between these galaxies and known X-ray sub-structures from \citet{neumann2003}, which indicates that stripping from the shocked ICM might be an important factor in triggering starbursts, and subsequent quenching, for infalling galaxies.

Studies of the entire supercluster population had to wait for new all-sky surveys with sufficient sensitivity to detect galaxies down to dwarf masses. Furthermore, a positive identification of supercluster members necessitates spectroscopic redshifts, which, prior to the SDSS catalog, only existed for the most massive galaxies and those immediately around the two clusters. \citet{kauffmann2004} were the first to conduct an extensive survey of the environmental dependence of star-formation activity in galaxies using the SDSS, by comparing the SFRs, among other spectroscopic and photometric measures of galaxy properties, to the local density around $\sim$122 000 galaxies in the SDSS Data Release One (DR1). They found that for galaxies at fixed stellar masses, the SFRs sharply decline at higher densities, and that the presence of an active galactic nucleus (AGN) is also much more common in galaxies with greater local density. \citet{haines2007} used $\sim3\times10^4$ low-$z$ galaxies in SDSS DR4, and showed a strong bimodal distribution between active and passive SF activity, using the equivalent width EW(H$\alpha$), and determined that the passive galaxies preferentially lie in regions with higher local galaxy density. One of the first major systematic studies of the environmental dependence of galaxy properties across Coma, using spectroscopically-confirmed members, was by \citet{gavazzi2010}. They used SDSS DR7 to select $\sim$4000 supercluster members, and characterised the local environment around each galaxy by measuring the volume density of galaxies within a cylinder of radius $1h^{-1}$Mpc and a half-length of 1000 km s$^{-1}$ (but with a slightly modified treatment of galaxies associated with the clusters, as described in Appendix \ref{appendix_b}). They examined the optical colours, morphologies, and frequency of post-starburst galaxies in different environments, and found a weak dependence of galaxy colour and morphology with environment for the most massive galaxies and a strong dependence of colour and morphology on environment for dwarf galaxies, and also that almost all post-starburst galaxies reside in higher-density regions. 

\citet{mahajan2010}, using a similar sample of SDSS-selected members of Coma, examined the local fraction of star-forming and AGN-hosting galaxies (characterised using SDSS spectral line measurements) vs local projected density. They found a similar broad trend as \citet{gavazzi2010}, whereby star formation in dwarf galaxies is strongly quenched at higher densities, and more massive galaxies show a weaker dependence on local density. \citet{mahajan2010} also used observations from \emph{Spitzer} MIPS at 24$\mu$$m$, which are available only for the core regions of A1656 and A1367, to obtain a complete accounting of star formation activity in the highest-density regions of Coma. For A1656, the more massive of the two clusters, they found significant IR detections only in the infalling regions, and recovered the expected correlation between dust-obscured and unobscured SFR with cluster-centric radius. However, for the less massive A1367 they found the reverse radial dependence for the fraction of star-forming galaxies when SF activity is derived by the optical measure vs IR measure. This result seems to indicate that an accounting of all tracers of SF activity, un-obscured and dust-obscured, may be required to get a clear picture of the quenching of SFR in galaxies.

With the release of the WISE all-sky survey, we now have access to the IR component of SF activity throughout the entire Coma Supercluster, and so this work presents the first look sensitive to un-obscured and dust obscured SF activity for virtually all galaxies in all environments of the Coma Supercluster, down to quiescent SFRs and with a sample of galaxies spanning over two orders of magnitude in stellar mass. With our SFR sensitivity, and our techniques for mapping the components of the cosmic web, we are in a position to put significant quantitative constraints on the degree to which pre-processing affects galaxies at $z\sim0$.

\subsection{SDSS}\label{sdss}
\begin{figure}
	\includegraphics[width=3.4in]{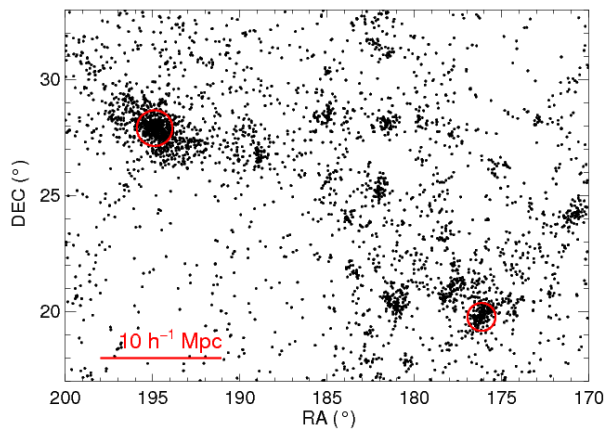}
 \caption{The Coma Supercluster, with member galaxies selected from the SDSS as described in Section \ref{sdss}. The two massive clusters (NE: A1656, SW: A1367) are identified by the red circles, which have radii equal to their virial radii. The virial radii come from $R_{200}$ measurements in \citet{rines2003}}.
	\label{fig:coma_galaxies}
\end{figure}

Our sample of supercluster galaxies is selected from DR9 of the Sloan Digital Sky Survey \citep[SDSS;][]{york2000}, which has mapped a large fraction of the sky in $ugriz$ bands and performed an extensive optical spectroscopic campaign complete (over the SDSS coverage areas) for galaxies with $r < 17.77$ mag. We select Coma Supercluster members from the SDSS DR9 galaxy sample following the selection criteria used by \citet{mahajan2010}, choosing galaxies with positions ($170\degr \le RA (J2000) \le 200\degr$, $17\degr \le DEC (J2000) \le 33\degr$) consistent with the Coma Supercluster and line-of-sight velocities, $cz$, within $\pm$2000 km s$^{-1}$ of either A1656 ($cz=6973$ km s$^{-1}$) or A1367 ($cz=6495$ km s$^{-1}$), where the central velocity of each cluster comes from \citet{rines2003}. To ensure no duplicate objects in our sample, we use only galaxies with the SPECPRIMARY designation set. To be sure that our sample is spectroscopically complete over the whole supercluster, we select only galaxies with $r\leq$17.77 mag. We have also excluded any galaxies with a ZWARNING flag to indicate a poor redshift determination (which affects less than one per cent of galaxies in the sample). We also require detections in the WISE 3.4 and 4.5$\mu$m bands for our entire sample (less than 1.5 per cent of our sample lacks detections in these two bands), and we apply a stellar mass cut-off (using stellar masses calculated with WISE photometry, see Section \ref{wise}) such that all galaxies in our sample have M$_*\ge10^{8.5}\msun$. The M$_*$ cut-off excludes just 130 galaxies, but it's necessary to prevent our sample, which is $r$-band selected, from being biased at the lowest galaxy masses towards only those dwarf galaxies which are the most actively star-forming. These selection criteria result in a sample of 3505 galaxies over the supercluster region covering $\sim$500 deg$^{2}$ on the sky. Figure \ref{fig:coma_galaxies} plots the galaxy positions over the supercluster. The virial radii for the two clusters plotted in Figure \ref{fig:coma_galaxies} come from the $R_{200}$ (the radius within which the density is equal to 200 times that of the critical density) values determined by \citet{rines2003}.

Having SDSS spectra for all of our galaxies, we can also mitigate the contributions of galaxies dominated by an AGN, which can otherwise contaminate our SFR estimates. \citet{brinchmann2004} used the emission lines of the SDSS galaxy spectra to classify galaxies according to a Baldwin, Phillips, \& Terlevich (BPT) diagram \citep{baldwin1981}, which cleanly delineates those galaxies which host LINER and AGN emission compared to those dominated by emission from HII regions. We use the \citet{brinchmann2004} classifications to identify galaxies dominated by AGN or LINER emission, and we exclude the WISE 22$\mu$m observations from these galaxies when measuring their SF activity throughout. Another benefit of having SDSS spectral information on our galaxies is access to the `4000\AA\ break' index, $D_n4000$, which is a measurement of the ratio of the average flux density in two narrow continuum bands, 3850-3950\AA\ and 4000-4100\AA\ \citep{balogh1999}. This index correlates strongly with the age of the stellar population in a galaxy, and has been shown to be a robust proxy for separating quiescent `red sequence' galaxies from those that are more actively star-forming in the `blue cloud', with the approximate dividing line between these galaxy populations at $D_n4000\sim1.6$ \citep{treyer2007, wyder2007}. Hereafter, we use measurements of $D_n4000$ for SDSS galaxies from \citet{kauffmann2003}. We also make use of the H$\delta$ line, specifically the index H$\delta_A$ \citep{wortheyottaviani1997, kauffmann2003} which is described in greater detail in Section \ref{kpa_vs_environ}.

We correct the optical-to-NIR band photometry for reddening using the \citet{schlegel1998} extinction maps, assuming the extinction curve of \citet{cardelli1989} with R$_V$=3.1. For the {\it GALEX} bands we used the extinction corrections of \citet{wyder2007}. We applied \textit{k}-corrections to our photometry using kcorrect v4\_2 \citep{blantonroweis2007} to get all UV-through-NIR photometry into the rest frame.

\subsection{GALEX}\label{galex}
We measure the un-obscured component of SF activity in Coma Supercluster galaxies from the {\it Galaxy Evolution Explorer} \citep[GALEX;][]{martin2005} GR6/GR7 data release, which includes mappings of the supercluster in Near-UV (NUV; 1750-2750 \AA) and Far-UV (FUV; 1350-1750 \AA) bands. Matching the SDSS galaxy catalogue to GALEX was done using the Mikulski Archive for Space Telescopes (MAST) database, with a 4$\arcsec$ search radius centred on the SDSS galaxy positions. In cases of multiple GALEX matches within the search radius, the GALEX match with a position closest to that of the SDSS galaxy coordinate was used. The GALEX bands are most sensitive to the photospheric emission of stars with masses $\ge 5 M_{\odot}$, and thus the UV continuum measurements provide an excellent tracer of recent star formation. Under the assumption of a star formation timescale that's long relative to the ages of these massive stars ($\tau_{SF} > 10^8$ yr), and a chosen IMF, one can derive a SFR$_{UV}$ corresponding to a given L$_{UV}$. The conversion between an observed luminosity and a SFR will be accurate as long as the emission picked up in one's UV band is dominated by the light of stars younger than 10$^8$ years. Although the FUV and NUV bands are both dominated by emission from young stars, if there is recent or on-going SF activity, the NUV band contains a greater fraction of contaminating flux from stars as old as $\sim$10$^9$ years \citep{hao2011, johnson2013}. 

\begin{figure}
	\includegraphics[width=3.2in]{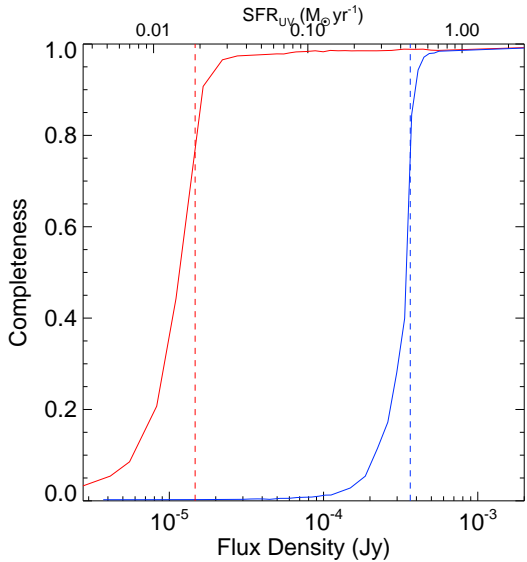}
	\caption{Completeness as a function of flux density for the shallowest {\it GALEX} mappings of the Coma Supercluster. In blue we plot the completeness in the FUV band, and in red the NUV. The two vertical dashed lines indicate the flux corresponding to 75\% completeness, with the upper x-axis denoting the corresponding SFR$_{UV}$ at the mean redshift of our Coma Supercluster sample.}
	\label{fig:galex_completeness}
\end{figure}

Our survey covers roughly 500 sq.~degrees, and so it is unsurprising that the {\it GALEX} observation depths vary greatly across the supercluster. Almost the entire supercluster has been mapped with {\it GALEX} at various depths, but about 5 per cent of our galaxy sample of supercluster members do not lie in a {\it GALEX} coverage area. The galaxies that are outside of {\it GALEX} coverage regions are flagged so that they are excluded from further analysis involving SF activity in the supercluster, as the SFRs we measure from WISE alone will necessarily be lower limits. The shallowest observation with {\it GALEX} in Coma has an exposure time of just 60 seconds, while the deepest is about 3$\times$10$^{4}$ seconds. Therefore, to ensure that the sensitivity of our catalogue to SFR$_{UV}$ is uniform across the supercluster, we must carefully account for the variation in completeness due to differences in survey depth. We measure completeness in a representative sample of {\it GALEX} NUV and FUV maps in Coma, including the shallowest maps, by extracting a supercluster galaxy detected in the map and re-inserting that galaxy, with a range of normalisations, into the maps. For each normalisation, we insert 100 of these `fake' galaxies into a FUV and NUV map with random positions, and then repeat 100 times for a total of 10$^4$ randomly placed galaxies per flux bin. We then run Source Extractor \citep{bertinarnouts1996} on each of the 100 maps per flux bin to determine the fraction of the `fake' galaxies that we recover as a function of flux density. Figure \ref{fig:galex_completeness} shows the completeness that we measure for the shallowest FUV and NUV map (with 60 second exposure time) of the Coma Supercluster. We have taken the fluxes corresponding to 75 per cent completeness in the shallowest FUV and NUV maps (indicated by vertical dashed lines in Figure \ref{fig:galex_completeness}), and we exclude any UV data for galaxies detected with fluxes below these completeness thresholds from our results. The completeness limit in NUV indicates that our Coma Supercluster catalogue is 75\% complete to SFR$_{UV}\ga0.02\sfr$ in all environments. 

As Figure \ref{fig:galex_completeness} shows, our SFR sensitivity in the Coma Supercluster is far greater in the NUV than in the FUV, and therefore we choose to use the NUV band to derive SFRs. To avoid having our SFR estimates significantly skewed by the presence of an older stellar population, which can contaminate the NUV band to a greater degree than in the FUV, we will ignore NUV-based SFR estimates for any galaxy whose SDSS spectrum shows a strong `4000 \AA\ break', based on $D_n4000>1.6$. This is explored in detail in section \ref{galexsfrs}.

\subsection{WISE}\label{wise}
A complete measure of SFR, especially for galaxies with reasonably high dust content, must include dust-obscured (indirect) tracers of SF activity. A significant step forward in the measurement of star-formation activity in the Coma Supercluster has recently been made possible with the all-sky data release from the Wide-field Infrared Survey Explorer \citep[WISE;][]{wright2010}, which has mapped the mid-infrared sky in four bands (W1-W4) centred at 3.4, 4.6, 12.0, and 22.0$\mu$m. Of particular interest to our analysis is the fourth band, which probes the blue-ward side of the dust emission curve in star-forming galaxies. We match our SDSS catalogue to the WISE point source catalogue by searching in a 5$\arcsec$ radius around each SDSS galaxy position, and selecting the WISE match whose position is closest to that of the SDSS galaxy. For galaxies matched to the WISE point source catalogue, we ignore W4 fluxes whose signal-to-noise in W4 is less than three. We measured completeness in W4 for a representative sample of the Coma Supercluster, following the procedure outlined in Section \ref{galex}. For WISE W4 we are 75\% complete to f$_{[22]}$ $\ge$ 4.7 mJy, which means our measurements of SFR$_{IR}$ are complete to SFR$_{IR}\ga$0.2 $\sfr$ for members of the Coma Supercluster (at L$_{IR} \ge 2.1\times10^8 \lsun$). For comparison, the only survey prior to WISE capable of detecting dust-obscured star formation activity across the entire supercluster was IRAS \citep{neugebauer1984}, whose completeness limit for L$_{IR}$ at Coma with its 25$\mu$m band is a full two orders of magnitude higher (L$_{IR} \ge 5\times10^{10} \lsun$). 

\begin{figure}
	\includegraphics[width=3.1in]{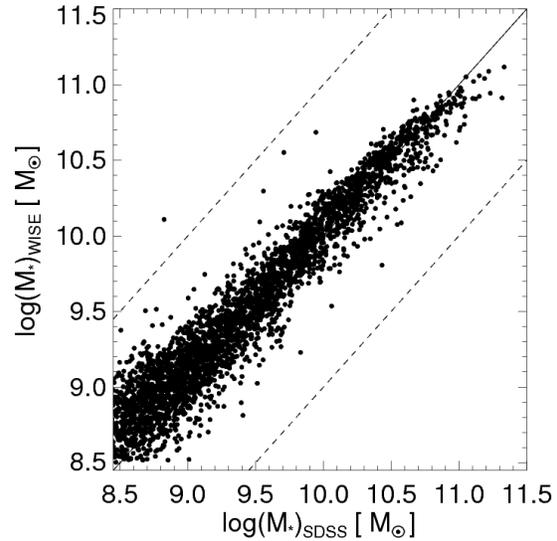}
	\caption{Comparison of stellar mass estimates from the WISE bands W1 + W2 \citep[using the calibration of][converted to a Kroupa IMF]{eskew2012} versus estimates from \citet{kauffmann2003} (using fits of SDSS spectra to stellar population synthesis models, and also with a Kroupa IMF) for galaxies in the Coma Supercluster. The solid line denotes a 1:1 correlation, and the dashed lines denote the boundaries of a systematic over- or under-prediction of a factor of 10.}
	\label{fig:st_masses}
\end{figure}

The two shortest-wavelength WISE bands, which sample the red side of the 1.6 \micron\ stellar photospheric feature, can be used to robustly estimate total galactic stellar masses (M$_*$). Recent work by \citet*{eskew2012} gives a calibration between flux densities measured in \emph{Spitzer} IRAC ch1 and ch2 and a galaxy's stellar mass (assuming a Salpeter IMF). We have applied the \citet{eskew2012} calibration, converted to a Kroupa IMF to be consistent with the rest of our study, using the WISE bands W1 and W2. In Figure \ref{fig:st_masses} we plot a comparison between the stellar masses of Coma Supercluster galaxies using the \citet{eskew2012} calibration with W1 + W2 photometry and the stellar mass estimates from the SDSS \citep{kauffmann2003}, where we find excellent agreement between these two independent measures of stellar mass.

\section{Mapping the Supercluster Environment}\label{lss}
A key component of our analysis is to characterise the local environment in a physically meaningful way. Typically, this approach has involved a calculation of the local surface (2D) or volume (3D) density of galaxies to describe the environment near a given galaxy based on the local density. A proper characterization of environment requires not just a measurement of the local galaxy density, but a technique to resolve the structures (groups, filaments, etc.) traced by galaxies. The latter task can be very difficult, as LSS mapping techniques are often susceptible to biases introduced by a characteristic shape or size scale one is examining. We have developed a technique to map LSS and characterise galaxy environments in a manner which is independent of the shape and size scale of the structure, and allows one to estimate local projected galaxy density over an arbitrarily large dynamic range of densities, by using a combination of Voronoi Tessellation (to calculate local surface density) and the Minimal Spanning Tree (to resolve continuous structures).

\subsection{Voronoi Tessellation}\label{vt}
VT is a method of decomposing a set of points into polygonal cells (Voronoi cells), where each cell corresponds to one point and the boundaries of the cell enclose all of the surrounding space closest to that point. VT has proven to be a very powerful tool for characterizing the galaxy density field \citep[e.g.,][]{platen2011, scoville2013}. When using a 2D surface density measurement to define the local galaxy density, one must be careful to not allow projection effects to significantly contaminate the density estimates. To help mitigate this we are using only spectroscopically-confirmed members. We show in Appendix \ref{appendix_b} that our surface densities in the Coma Supercluster correlate strongly with volume density estimates calculated following the procedures of \citet{gavazzi2010}. For an additional strong demonstration of the effectiveness of surface density measurements as tracers of the volume density, see Figure 1 of \citet{gallazzi2009}, which shows a tight correlation between surface density and volume density over about two orders of magnitude in density being probed in the Abell 901/902 supercluster.

A common challenge when trying to measure the galaxy density field is determining the area (or volume) over which to measure the density at each galaxy's position. Often the density measurements are made on a size scale set by the n$^{th}$ nearest neighbour or by using a characteristic kernel with an adaptive size scale. Although these methods have flexible size scales, the shape of the region used to calculate the density field is generally fixed. Calculating densities reliably over a very large dynamic range of environments requires a technique that can adjust to arbitrary size scales \emph{and} local geometry. VT addresses the difficulty of needing both adaptive size and shape by using cells which automatically adjust to the nearby density, and which assume no {\it a priori} shape. In regions of lower source density the cells are larger on average, and the cells get progressively smaller in higher density regions. With VT, one can calculate the local density around a given galaxy's position by taking the inverse of the area of the cell that encloses that galaxy.

\begin{figure*}
\centering
\begin{tabular}{cc}
\includegraphics[width=3.8in]{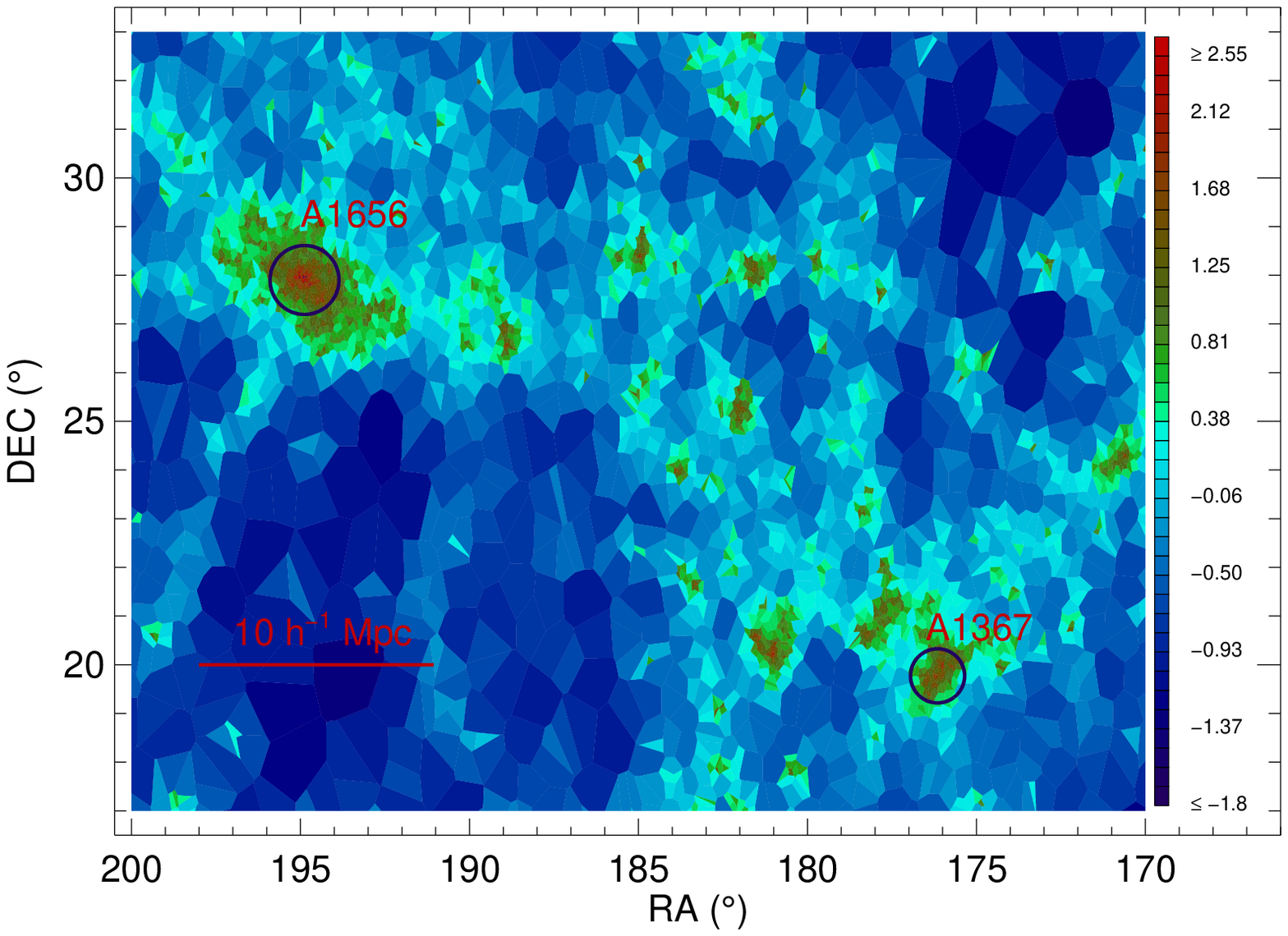} 
\includegraphics[width=2.75in]{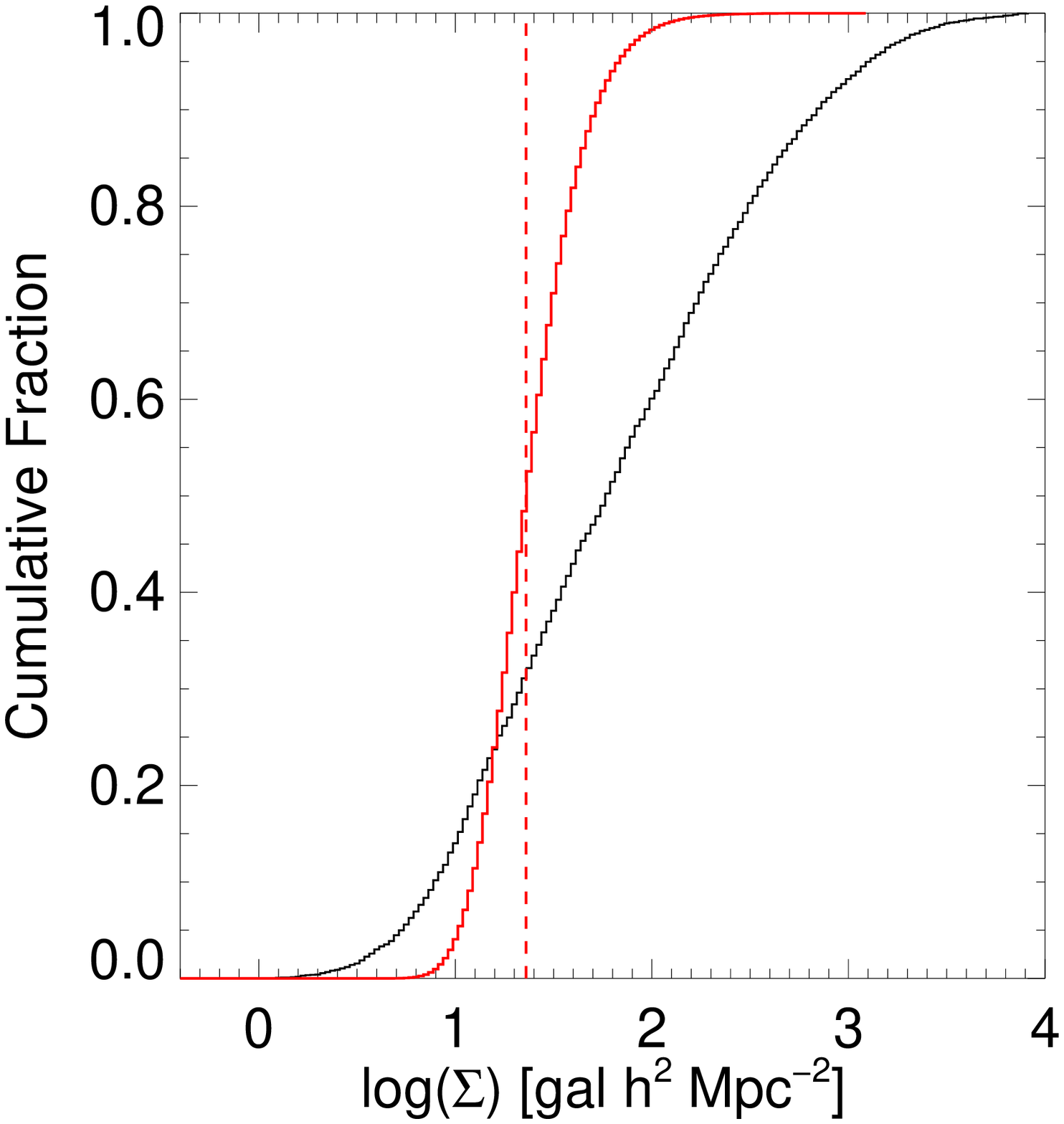}
\end{tabular}
\caption{(Left) VT of the Coma Supercluster, with cells colour coded by the density parameter $\delta$, derived from the figure to the right. (Right) Cumulative distributions of Voronoi cell projected densities in Coma (black solid line). We also plot the mean cumulative distribution of Voronoi cell surface densities for an ensemble of random maps generated with the same number of sources and same area as our Coma supercluster sample (red solid line). The median density of the ensemble of random maps is denoted by the vertical dashed line. The density indicated by the dashed line serves as the baseline for measuring $\delta$ for the left figure.}
\label{fig:tessduo}
\end{figure*}

We compute the VT of the Coma Supercluster using the QHULL function \citep{barber1996} in \textbf{IDL}, which calculates convex hulls for the 2D distribution of points. A common issue with VT, and with any method of measuring the local density field, is spurious density estimates arising near the edges of the map, where many cells can be artificially large or even unbounded. We avoid this issue entirely by adding galaxies from the SDSS DR9 in a 10 degree-wide `buffer' surrounding our Coma Supercluster map, which all have redshifts consistent with Coma and the same selection criteria defined in Section \ref{sdss}, when calculating our VT. We then exclude these buffer galaxies from further analysis, so the only purpose of these galaxies is to ensure that we do not suffer any edge effects in our supercluster dataset. In Figure \ref{fig:tessduo} (Left) we plot the Voronoi cells over the supercluster. We compare the distribution of cell densities in the supercluster to a set of 1500 maps generated with source positions randomly distributed, but each with an equal number of sources and an area equal to the Coma Supercluster map. In Figure \ref{fig:tessduo} (Right) we show the cumulative distribution of cell densities observed in Coma compared to the mean cumulative distribution of projected Voronoi cell densities for the set of random realizations. These random maps are necessary to establish a baseline density with which to compare our observed Voronoi cell densities in Coma.

Figure \ref{fig:tessduo} (Right) highlights a couple of key differences between the cell density distribution of the observed supercluster population and that of the random realizations. As one would expect, since the supercluster contains regions of extremely high galaxy density, there is a much larger fraction of cells with densities upwards of 100-1000 gal h$^2$ Mpc$^{-2}$ than in the random distributions. But we also find that there is a larger fraction of cells in the Coma Supercluster at very low densities, around 1-10 gal h$^2$ Mpc$^{-2}$, as a more clustered population implies that one finds more prominent voids as well. The cumulative distribution of projected cell densities in our full Coma catalogue gradually increases over a density range of log($\Sigma$)=0.5--3.5 [gal h$^2$ Mpc$^{-2}$], meaning that our supercluster map samples about three orders of magnitude in projected galaxy density. The random distributions tend to sample about one order of magnitude in galaxy density with any appreciable number of cells. VT measures the density field across the huge dynamic density range of the supercluster, but it is less effective at resolving continuous structures. Therefore, we now turn to the complementary approach of the MST.

\subsection{The Minimal Spanning Tree}\label{mst}
The MST is a technique for examining the local clustering properties of a distribution of points. The technique was first used for astronomical data analysis by \citet{barrow1985}, and was instrumental for the first statistically rigorous detection of a filamentary structure on cosmological scales, seen in the CfA catalogue \citep{bhavsarling1988}. The MST technique has been used fairly sporadically over the past couple of decades, but it has been put to use in a variety of astronomical contexts in recent years \citep[e.g.,][]{colberg2007, gutermuth2009, adami2010, durret2011}. Recently, a study of the Galaxy And Mass Assembly \citep[GAMA;][]{baldry2010, driver2011} fields by \citet{alpaslan2013} used the MST to trace filamentary structures by using the positions of galaxy groups identified previously in their fields.

\begin{figure*}
\centering
\begin{tabular}{cc}
\includegraphics[width=3.8in]{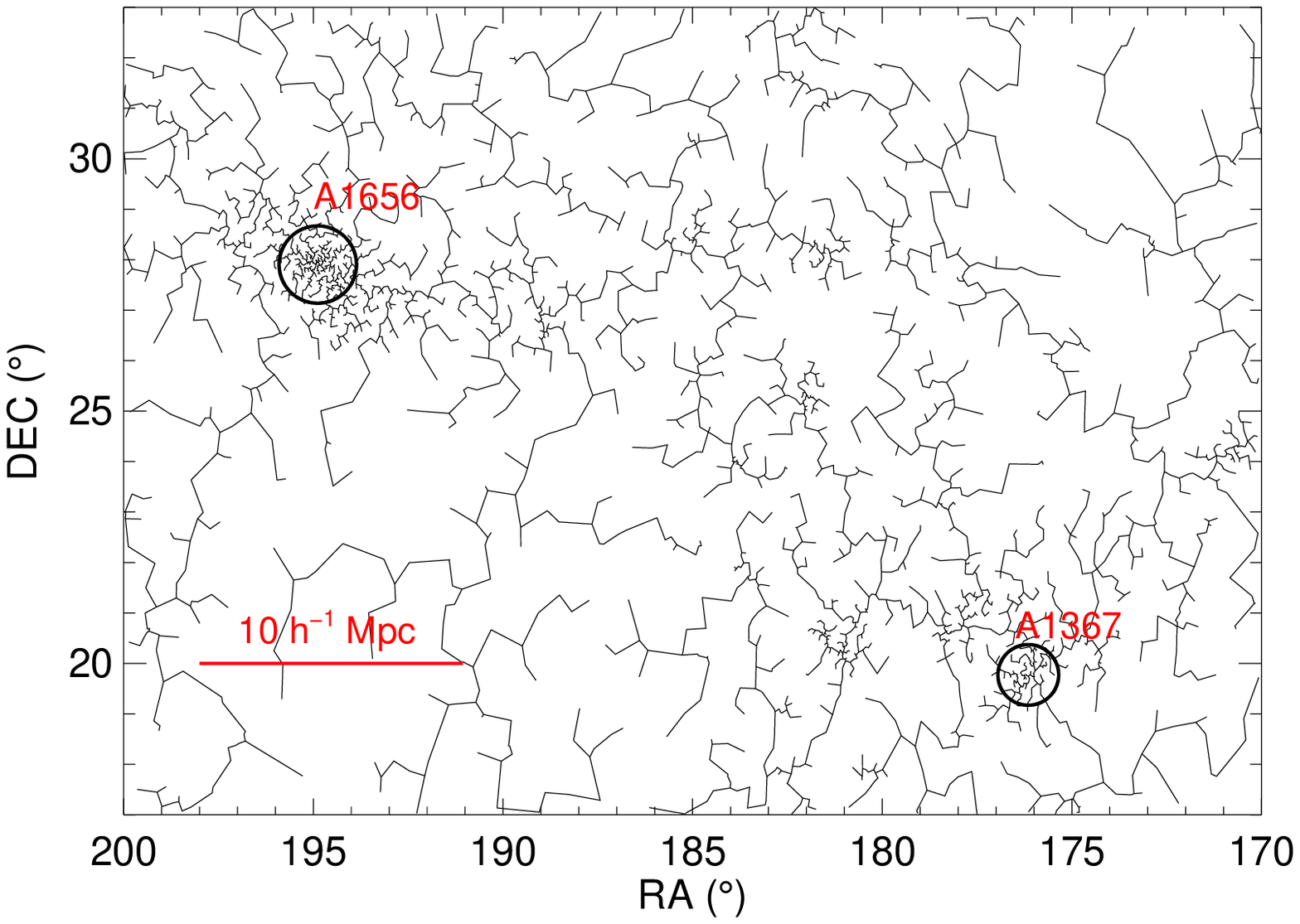} 
\includegraphics[width=2.75in]{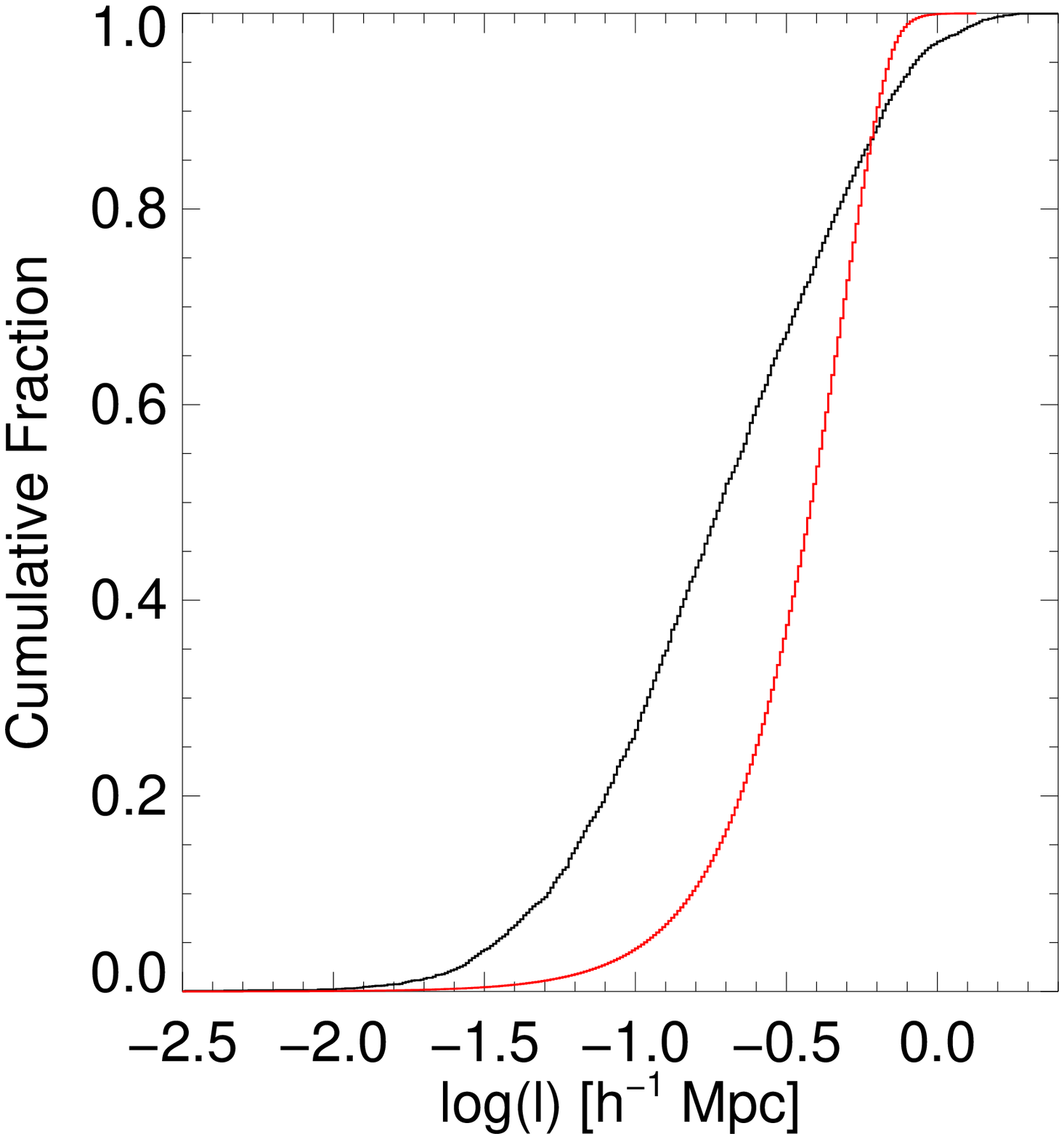}
\end{tabular}
\caption{(Left) MST of the Coma Supercluster, with the branches connecting galaxies drawn as solid lines. (Right) The cumulative distribution of MST branch lengths for the Coma Supercluster (black). Also plotted (in red) is the mean cumulative branch length distribution for the 1500 random maps.}
\label{fig:mstduo}
\end{figure*}

If we treat a distribution of galaxies as nodes, and connect all nodes with branches (straight lines) such that no two branches cross paths, then we have constructed a spanning tree. There are many possible configurations a spanning tree can manifest given a series of nodes, but a spanning tree whose total branch length is a minimum is a MST. To construct a MST from the distribution of galaxies in the Coma Supercluster, we use a custom \textbf{IDL} code written by R. Gutermuth \citep[see][]{gutermuth2009}. In order to extract structures from a MST we must select a critical branch length, $l_{crit}$, chosen so that subsets of the spanning tree with nodes connected entirely by branches of length $l\leq$$l_{crit}$ are separated (pruned, if you will) from the tree and considered a distinct structure of nodes. In this manner, we can identify continuous structures traced by the galaxy distribution, such as filaments, clusters, and groups, with any arbitrary shape, exploiting the changes in the local clustering properties associated with the angular power spectrum. There are two free parameters used to extract structures from a MST: $l_{crit}$ and the minimum number of nodes necessary in a structure (throughout our analysis we use a minimum number of eight galaxies to avoid any chance of projection effects leading to false positives). In Figure \ref{fig:mstduo} (Left) we show our MST of the Coma Supercluster. Figure \ref{fig:mstduo} (Right) gives the cumulative distribution of branch lengths in our MST of Coma, and also plots the mean cumulative distribution of branch lengths from a MST computed over the ensemble of random realizations described in Section \ref{vt}. Our cumulative branch length distribution fundamentally mirrors the cumulative distribution of Voronoi cell densities, with large branches corresponding to low cell densities, and vice-versa. 

In Figure \ref{fig:tessduo} we saw that the galaxies in our supercluster reside in an extremely wide range of local densities. We can therefore expect that the characteristic clustering scales of galaxies in the supercluster also vary greatly with environment.  To identify structures ranging from dense clusters to diffuse filaments and voids traced by the galaxy distribution we must apply a multi-tiered approach to our mapping of the environment with the MST, whereby multiple critical branch lengths are used to select structures of different characteristic densities. Our aim is to identify clusters, groups, filaments, and voids in the Coma Supercluster, and so we choose two particular critical branch lengths for our analysis, $l_{crit1}$ and $l_{crit2}$; the former branch length is chosen to delineate the boundary between clusters/groups and the filaments, while the latter is selected to represent the boundary between the filaments and voids. 

To choose $l_{crit1}$, we consider that groups and clusters should have a minimum projected density of $\sim$40 galaxies h$^2$ Mpc$^{-2}$ (this would vary depending on the depth of one's survey, but in our case it's a reasonable estimate). Then we note that the projected density of 40 galaxies h$^2$ Mpc$^{-2}$ corresponds to a cumulative fraction of 0.41 in our Voronoi cell density distribution (see Figure \ref{fig:tessduo}). And since the MST branch length distribution mirrors the Voronoi cell density distribution, the approximate branch length that would correspond to the density threshold of 40 galaxies h$^2$ Mpc$^{-2}$ can be found at a MST branch cumulative fraction of 0.59. For our MST of the Coma Supercluster, the branch length at a cumulative fraction of 0.59 is 516$\arcsec$, or $\sim$0.25 Mpc at the distance of the Coma Supercluster. Therefore, we identify structures in the MST corresponding to clusters and groups by choosing sub-sets of the tree which are all connected by branches of length $l\leq$516$\arcsec$, and which consist of at least eight members. Any such structure whose central position is within the virial radius of A1656 or A1367 is labeled as part of a galaxy cluster, while any structure located beyond the virial radii of the two clusters is labeled as a distinct galaxy group. Applying these selection criteria, we identify 741 cluster galaxies and 716 group galaxies out of the 3505 galaxies in our supercluster sample. In practice, our method of identifying continuous structures via the MST is very similar to the Friends of Friends (FoF) algorithm \citep{huchrageller1982, gellerhuchra1983}, but with only galaxy surface densities being considered. To explicitly verify the similarities between the MST and FoF approaches, we applied the FoF algorithm to our Coma Supercluster sample using a linking length equal to $l_{crit1}$, and find the same cluster and group structures are identified as with the MST.

The population of galaxies residing in filaments in our map can then be selected as those in structures connected by branches of length 516$\arcsec$$<$$l$$\leq$$l_{crit2}$. The determination of $l_{crit2}$, as with $l_{crit1}$, is complicated by the fact that there exists no clearly-defined delineation between populations of filament and void galaxies. However, to define $l_{crit2}$ we follow an approach similar to what we described previously. We begin by choosing an approximate projected density threshold corresponding to the transition between filament and void galaxies, which in this case we have taken as 10 galaxies h$^2$ Mpc$^{-2}$. Then we identify the cumulative fraction of Voronoi cells in Coma corresponding to that density (0.13), and therefore take $l_{crit2}$ as the branch length corresponding to a cumulative fraction of 0.87 in our cumulative branch length distribution, which is 1286$\arcsec$, or $\sim$0.61 Mpc at Coma. Using a $l_{crit2}$ of 1286$\arcsec$, and a minimum of eight galaxies per structure, and excluding any galaxy already identified as part of a group or cluster, we find 1292 galaxies residing in filaments in the Coma Supercluster. Finally, we select void galaxies as anything which has not been identified as a member of a cluster, group, or filament. This is generally any galaxy whose separation from its nearest neighbour is greater than $l_{crit2}$, but it also can include groupings of fewer than eight galaxies separated by less than the aforementioned critical branch lengths. Overall, we select 735 void galaxies in our Coma Supercluster sample. 

\begin{figure*}
	\includegraphics[width=4.6in]{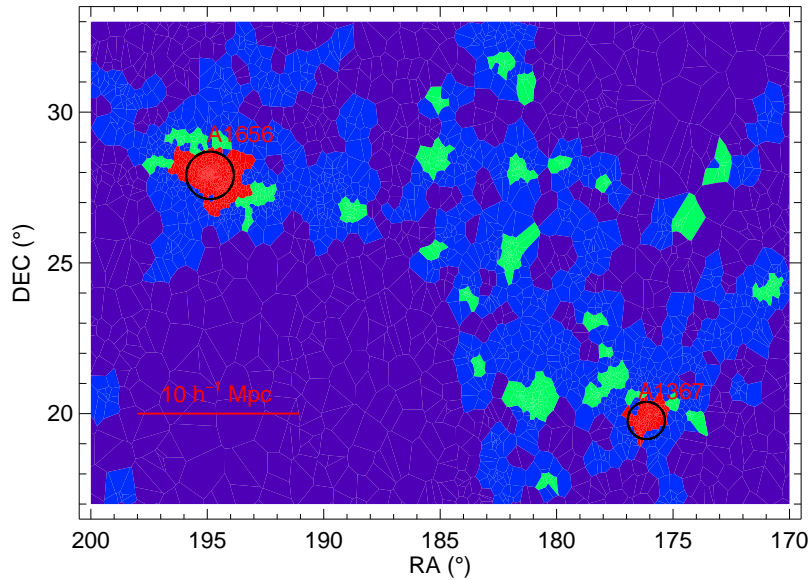}
	\caption{Coma Supercluster with Voronoi cells coloured by MST-defined environments, with cluster (red), group (green), filament (blue), and void (purple) galaxies selected as described in Section \ref{mst}.}
	\label{fig:mst_tiers}
\end{figure*}

\begin{figure*}
\centering
\begin{tabular}{cc}
\includegraphics[width=2.6in]{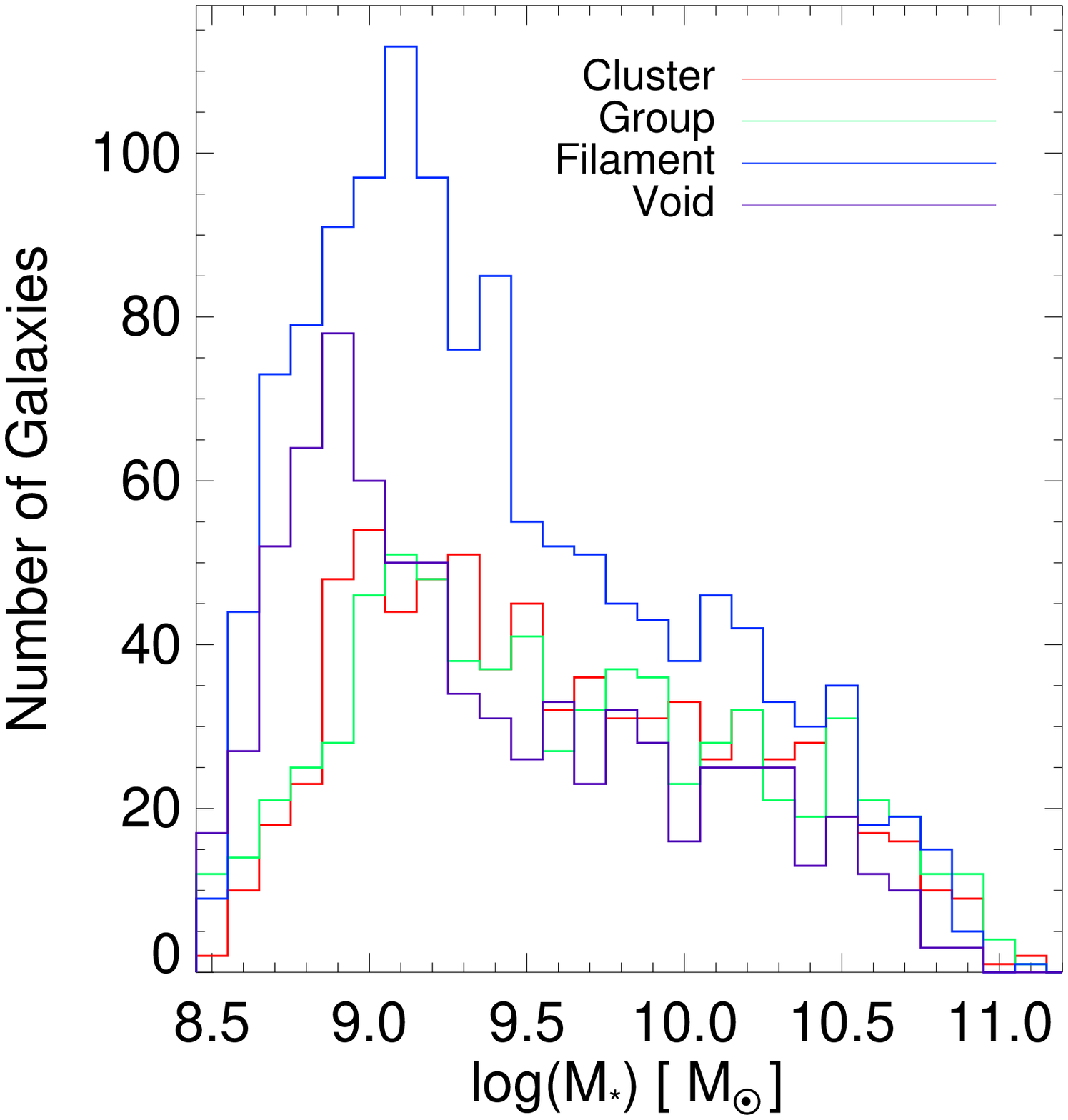} 
\includegraphics[width=2.6in]{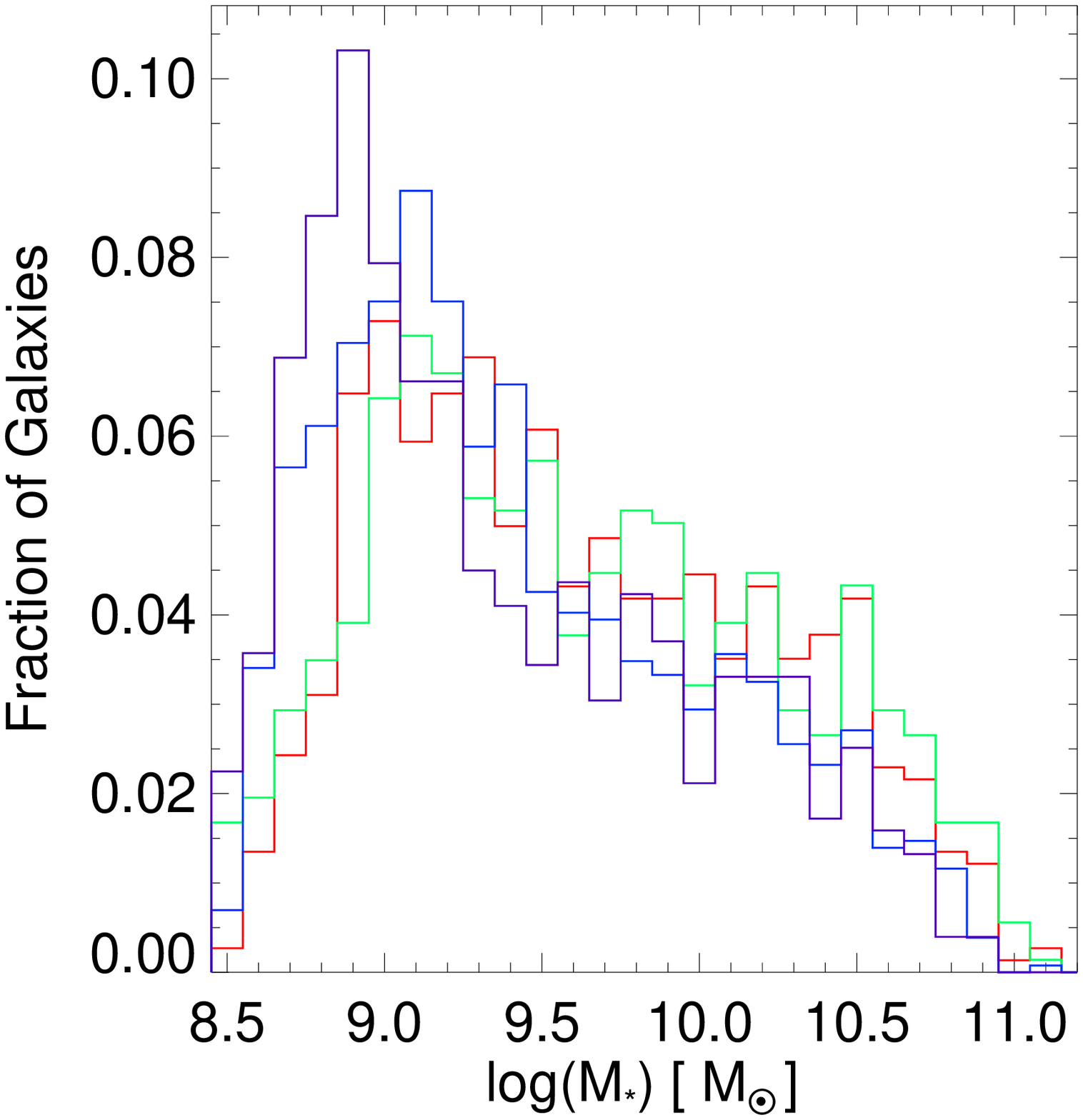}
\end{tabular}
\caption{(Left) The differential distribution of galaxy stellar masses for each of the four environments presented in Section \ref{mst}. (Right) The stellar mass distributions of each of the four environments plotted in terms of the fraction of galaxies within each environment. The denser environments are skewed towards more massive galaxies, while the less dense environments have a higher relative fraction of dwarf galaxies.}
\label{fig:stmass_distribution}
\end{figure*}

One reasonable concern about this technique is that it can produce environmental demographics that differ depending on the choices for characteristic densities (e.g., if we had used 50 galaxies h$^2$ Mpc$^{-2}$ to idenfity $l_{crit1}$ instead of 40, we would have fewer galaxies in the clusters and groups and more galaxies in the filaments), and therefore different conclusions might be drawn about the environmental dependence on galaxy evolution. We address this concern in Appendix \ref{appendix_a}, by showing that our results are insensitive even to large variations in the choices for $l_{crit}$. Figure \ref{fig:mst_tiers} shows a map of the Voronoi cells of the Coma Supercluster colour-coded to show cluster (red), group (green), filament (blue), and void (purple) galaxies. Our sample of 3505 Coma Supercluster galaxies is split such that we have approximately 20/20/40/20 per cent  in the cluster/group/filament/void environments. In Table \ref{tab:environs} we present the basic statistics of the four environments, including total surface areas, mean projected densities, and overall star-formation activity. These four environments also generally differ in terms of the stellar mass content of their constituent galaxies. Figure \ref{fig:stmass_distribution} presents the distribution of galaxy stellar masses in each of the four environments, which shows that the cluster and group environments are skewed towards higher-mass galaxies than the filament and void populations. When we run a Kolmogorov-Smirnov (KS) test comparing the stellar mass distributions of galaxies in each pair of environments, we find that only the stellar mass distributions of the cluster and group environments are consistent with being drawn from the same distribution (p$_{KS}=0.85$ for cluster-group). Running a KS test comparing all pairs of environments except for cluster-group yields p$_{KS}\leq1\times10^{-4}$, indicating statistically distinct stellar mass distributions.

\begin{table}
 \centering
  \caption{Environments of the Coma Supercluster defined using the MST technique, as outlined in Section \ref{mst}. Column 3 gives the total area of the Voronoi cells of the galaxies in each environment. Column 4 gives the mean Voronoi cell surface density of the galaxies in each environment. Column 5 gives the log of the average specific SFR (SSFR=SFR/M$_*$) of galaxies in each environment, taken as the sum of the SFRs of the galaxies in each environment (excluding galaxies which are dominated by an AGN, see Section \ref{sfrs}, and those which are not in \textit{GALEX} coverage areas) divided by the sum of the stellar mass of all galaxies in that environment (excluding the galaxies not in \textit{GALEX} coverage areas).} \label{environs}
  \begin{tabular}{@{}llllr@{}}
  \hline
   Environ.  &  N$_{gal}$  &  Area  &   $<$$\Sigma$$>$  &  log$(<$SSFR$>)$  \\
                &             &   (h$^{-2}$ Mpc$^{2}$) &  (h$^{2}$ Mpc$^{-2}$) &    [yr$^{-1}$]  \\
 \hline
 Cluster & 741  & 19.5  & 108.5 & -11.26  \\
 Group   & 716  & 57.5  & 38.7  & -11.06  \\
 Filament& 1292 & 423.6 & 6.3   & -10.60  \\
 Void    & 756  & 891.0 & 1.7   & -10.42  \\
\hline
\label{tab:environs}
\end{tabular}
\end{table}

\section{Results}\label{results}

\subsection{SFRs}\label{sfrs}

\subsubsection{GALEX}\label{galexsfrs}

As previously indicated in Section \ref{galex}, the UV coverage across the supercluster varies widely, and our completeness limits in the areas with the shallowest coverage do not afford us great SFR sensitivity in the FUV band, but the depths of the observations are much more favorable for the NUV. Measurements of SF activity in the UV with {\it GALEX} typically utilise the FUV band rather than the longer-wavelength NUV, because the NUV is known to suffer greater contamination by flux from older stellar populations with ages $\tau\sim$200Myr \citep{hao2011}. However, \citet{johnson2013} find that the degree to which the NUV luminosity of a galaxy is contaminated by an older stellar population strongly correlates with the star-formation history (SFH) of the galaxy. Therefore, we use the $D_n4000$ measurements from the SDSS spectra to directly examine the effect of SFH on the SFRs calculated with FUV and NUV. To get SFRs for FUV and NUV we use the following equations, both of which are derived from the \citet{kennicutt1998} calibration for a Kroupa IMF:

\begin{equation}\label{eq:galexsfrfuv}
	\frac{SFR_{FUV}}{\sfr}=4.42\times10^{-44}\left(\frac{L_{FUV}}{erg~s^{-1}}\right)
\end{equation}

\begin{equation}\label{eq:galexsfrnuv}
	\frac{SFR_{NUV}}{\sfr}=7.57\times10^{-44}\left(\frac{L_{NUV}}{erg~s^{-1}}\right).
\end{equation}

\begin{figure}
	\includegraphics[width=3.0in]{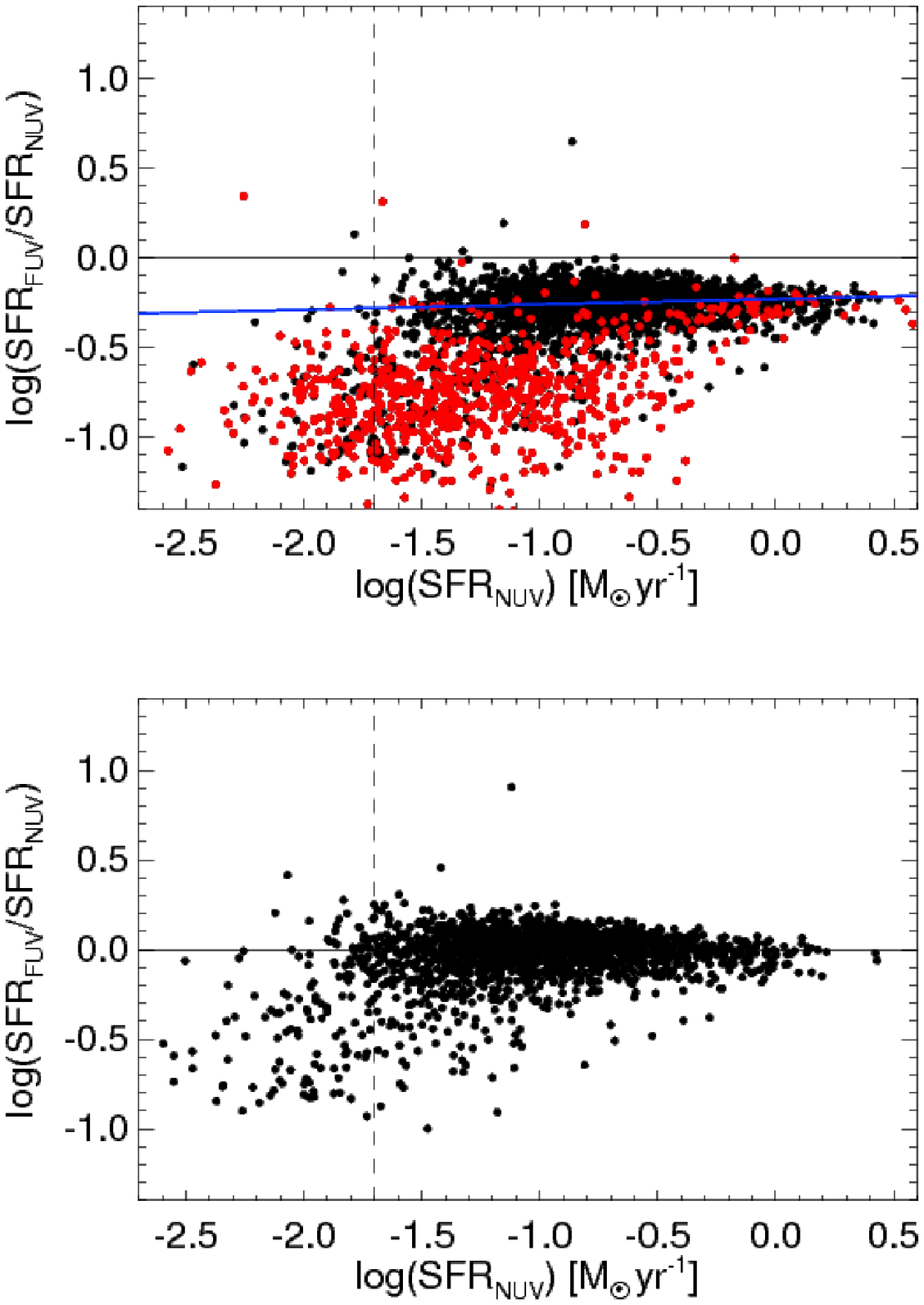}
	\caption{(Top) Comparison between SFRs calculated using Equations \ref{eq:galexsfrfuv} and \ref{eq:galexsfrnuv}. The black points are galaxies with $D_n4000<1.6$, while the red points have $D_n4000\ge1.6$. The vertical dashed line indicates the 75\% completeness threshold for SFR$_{NUV}$ in the shallowest coverage area for the supercluster. The blue line indicates a best-fitting to the distribution of galaxies with $D_n4000<1.4$, which are dominated by young stellar populations. (Bottom) The extinction-corrected SFR comparison for galaxies with $D_n4000<1.6$.}
	\label{fig:sfr_fuv_nuv}
\end{figure}

Note that there are two effects we must account for if we wish to use the NUV to reliably estimate SFRs. The first is the difference in the internal extinction of our galaxies in the FUV and NUV, and the second is the differing contaminations of older stellar populations on the SFR estimates. Both of these effects can be addressed by examining a comparison between SFR$_{FUV}$ and SFR$_{NUV}$, as shown in Figure \ref{fig:sfr_fuv_nuv}.

From Figure \ref{fig:sfr_fuv_nuv} (Top) we find that overall the SFRs calculated from Equations \ref{eq:galexsfrfuv} and \ref{eq:galexsfrnuv} show an over-prediction of the SFR$_{NUV}$ relative to SFR$_{FUV}$ because of a combination of the two effects described previously. The contamination due to older stellar populations clearly causes a very large (upwards of a factor of 10 in some cases) over-prediction, and with a large spread for galaxies with $D_n4000\ge1.6$. However, the galaxies with $D_n4000<1.6$ predominantly lie along an approximately flat distribution offset slightly from a 1:1 correlation with SFR$_{FUV}$. This slight offset exhibited by the galaxies dominated by younger stellar populations is caused by the difference in internal extinction in these galaxies between the FUV and NUV bands, and so we can bring the NUV-based SFRs in close agreement with those from the FUV by applying a correction to all L$_{NUV}$ calculations. Our NUV extinction-corrected SFRs therefore come from the equation \ref{eq:galexsfrnuv} after correcting L$_{NUV}$ for internal dust extinction. However, as the NUV clearly becomes unreliable as a SFR indicator when $D_n4000\ge1.6$, we only consider the UV component of SFRs for galaxies with $D_n4000<1.6$.

\subsubsection{WISE}\label{wisesfrs}

To calculate the infrared SFR from W4 luminosities, we use the calibrations devised for MIPS 24$\mu$m presented in \citet{murphy2011} [equation 5]:

\begin{equation}\label{eq:wisesfr}
	\frac{SFR_{22\mu m}}{\sfr}=5.58\times10^{-36}\left[\frac{\nu L_{\nu}(22\mu m)}{erg~s^{-1}}\right]^{0.826}.
\end{equation}

\citet{goto2011} conclusively established that the MIPS-SFR calibration is accurate with the W4 band with just a 4\% scatter, well within the typical uncertainties of SFR$_{IR}$ estimates. To verify the accuracy of our SFR$_{IR}$ measurements, we obtained 70$\mu$$m$ observations made with the PACS instrument \citep{poglitsch2010} on the Herschel Space Observatory \citep{pilbratt2010} over a subset of the Coma Supercluster. The observations we obtained (PI: C. Simpson) cover 1.75 sq.~degrees centered on A1656. We downloaded the level 2\_5 processed data products using the Herschel Interactive Processing Environment \citep[HIPE,][]{ott2010} tool, and extracted sources and fluxes within HIPE using the source extractor SUSSEXtractor \citep{savageoliver2007}. After matching the Herschel sources to our Coma Supercluster catalogue, we measured comparison SFR$_{IR}$ values using the calibration of \citet{calzetti2010}, and found that these comparison SFRs are in good agreement with the SFRs we calculate with Equation \ref{eq:wisesfr}.

\subsubsection{GALEX + WISE}\label{galexwisesfrs}

Utilising both the IR and UV measures of SFR, we can estimate the total bolometric SFR, assuming that SFR$_{tot}$ = SFR$_{IR}$ + SFR$_{UV}$, following equation 9 of \citet{murphy2011}, but for NUV rather than FUV:

\begin{equation}\label{eq:totsfr}
	\frac{SFR_{tot}}{\sfr}=7.57\times10^{-44}\left(\frac{L_{NUV} + 0.88L_{IR}}{erg~s^{-1}}\right).
\end{equation}

For any galaxy detected in WISE W4 but not in NUV (or if the NUV flux is below the completeness threshold described in Section \ref{galex}), we calculate its SFR using equation \ref{eq:wisesfr}, assuming that SFR$_{tot}\simeq$SFR$_{IR}$. Similarly, when a galaxy is detected in NUV but not in W4 we use equation \ref{eq:galexsfrnuv}, but in cases of detections in both NUV and W4 we use equation \ref{eq:totsfr}. 

Out of the 3505 galaxies in our Coma Supercluster sample, 1039 are detected in WISE W4 with a S/N of at least 3.0, but 131 of these galaxies have spectra indicative of a dominant AGN. Therefore, we have the IR contribution to SFRs measured for 908 galaxies. Although 3139 galaxies are detected in the NUV in our sample, only 2703 galaxies have a NUV flux above the completeness threshold shown in Figure \ref{fig:galex_completeness}. Therefore, we measure the UV component of SFR only for these 2703 galaxies. All together, we have SFR measurements for 2798 galaxies: 95 with SFRs measured only in WISE W4, 1890 with SFRs measured only in NUV, and 813 whose SFRs are derived from a combination of NUV and W4.

\subsection{SFR versus Environment}\label{sfr_vs_environ}

Next we examine the SFRs of Coma Supercluster galaxies as a function of environment using two complementary approaches: by calculating the fraction of galaxies in each environment that are star-forming and by examining the specific SFR (SSFR=SFR/M$_*$) and SFR distributions of star-forming galaxies in each environment. Hereafter we define galaxies as `star-forming' (SF) if they have log(SSFR)$\geq$-11[yr$^{-1}$] and, to avoid potential over-estimation of SFR from contamination of the NUV by an older stellar population, $D_n4000<1.6$. Our choice for SSFR threshold was motivated by our observation, as in other studies \citep*[e.g.,][]{wetzel2012}, that the galaxy population shows distinct bimodality about SSFR$\simeq$10$^{-11}$ yr$^{-1}$. Adjusting the threshold by which we define a galaxy as SF, e.g. by using the \citet{elbaz2011} `star-forming main sequence', systematically shifts the fractions of SF galaxies in each environment but does not affect the overall trends of SF activity versus environment in our study.

\subsubsection{SF Fraction versus Environment}\label{sf_fract_vs_environ}
\begin{figure*}
\centering
\begin{tabular}{cc}
\includegraphics[width=3.4in]{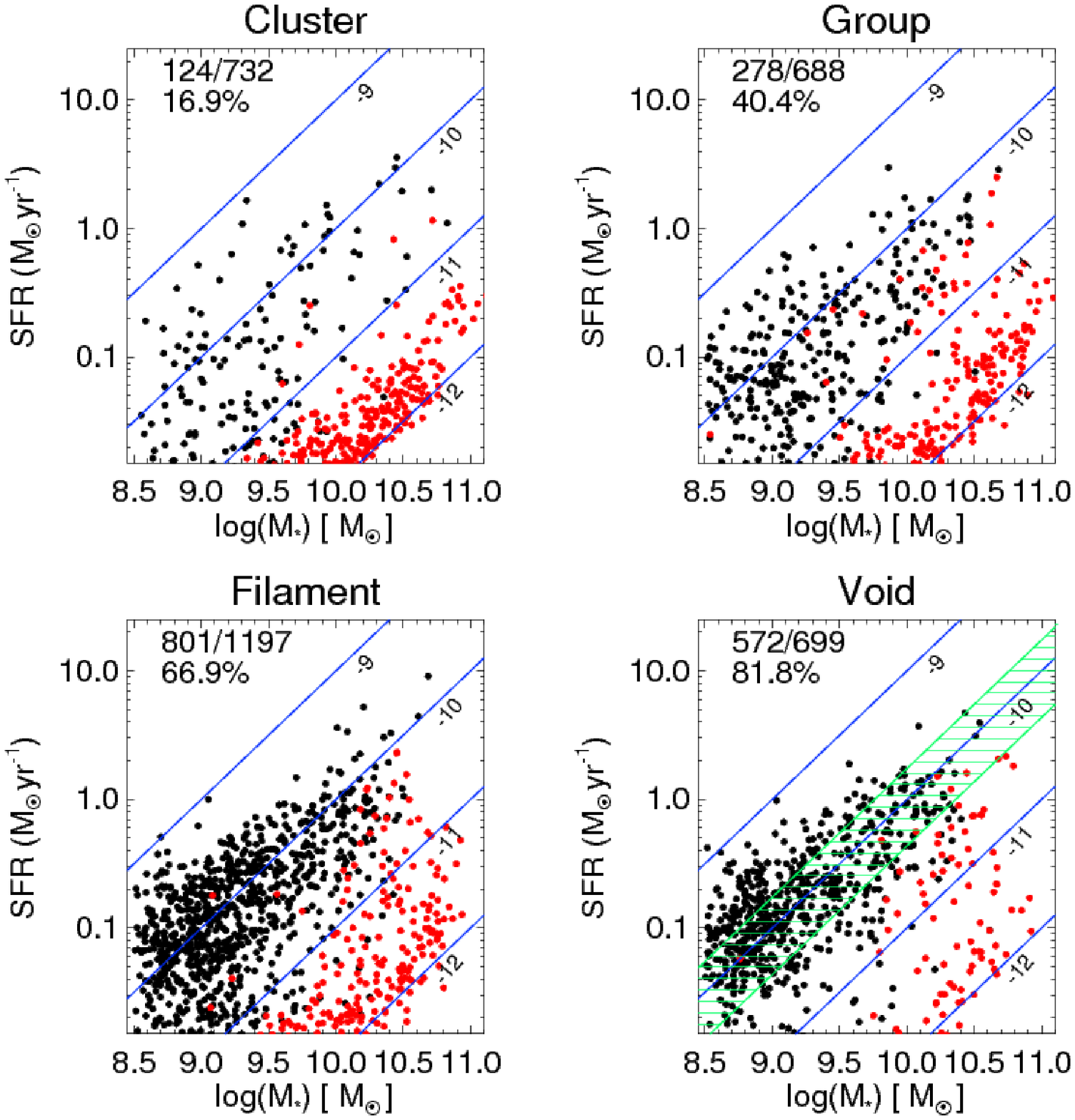}
\includegraphics[width=3.4in]{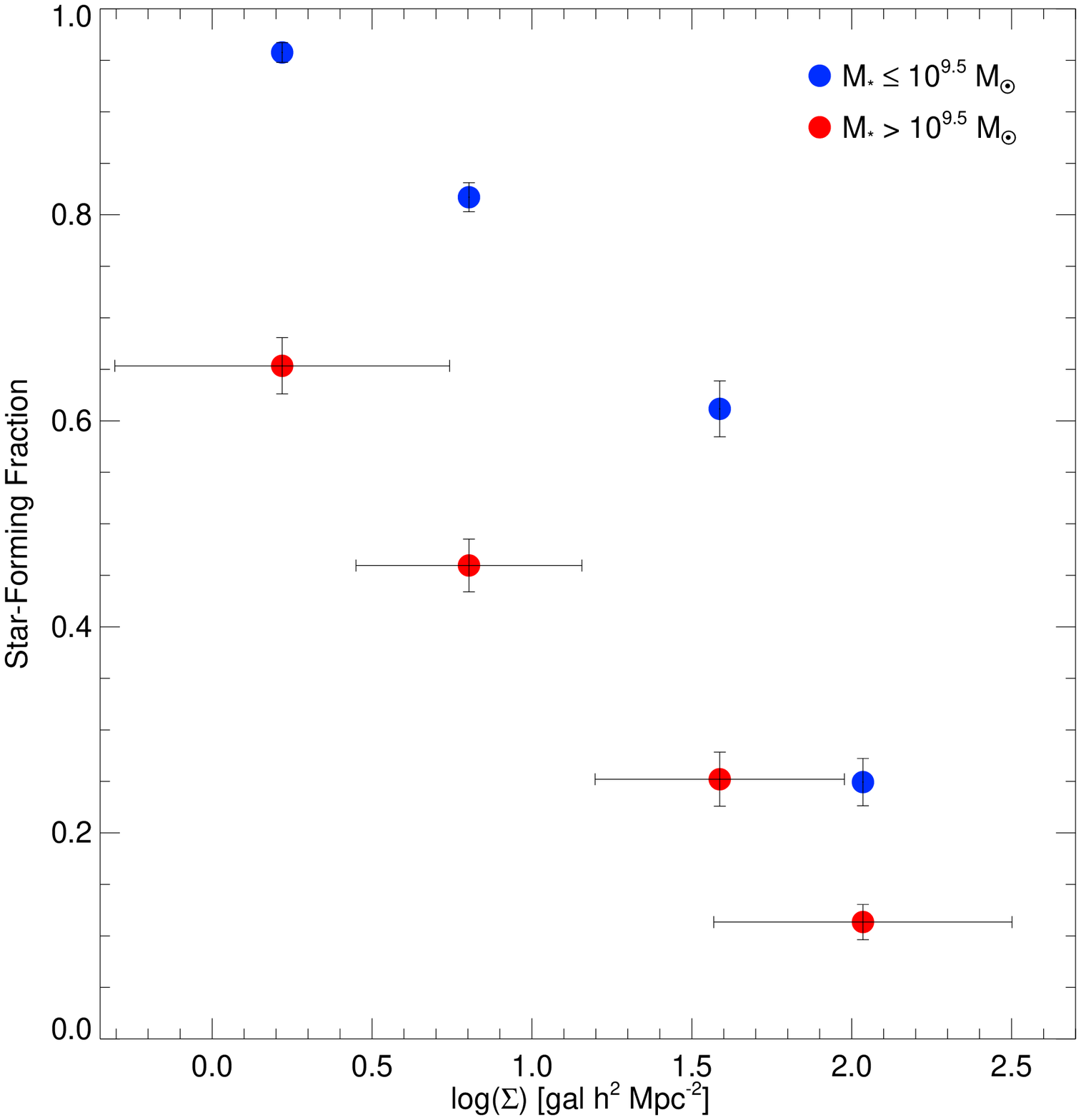}
\end{tabular}
\caption{(Left) SFRs vs stellar mass for galaxies in the four environments described in Section \ref{mst}. The red points correspond to galaxies with $D_n4000\ge1.6$, for which we expect the SFR to be over-predicted based on L$_{NUV}$ (see Section \ref{galexsfrs}). The solid blue lines indicate constant specific SFR, with labels to denote the log(SSFR) [yr$^{-1}$]. The fractions in the upper left corner of each panel indicate the ratio of galaxies with SFRs above the completeness threshold (and $D_n4000<1.6$) over the total number of galaxies in each environment, and the ratios are also expressed as a percentage. The void galaxy plot also includes a green shaded region, which indicates the star-forming `main sequence' for galaxies at the redshift of the Coma Supercluster, as defined by \citet{elbaz2011}. (Right) Fraction of SF galaxies, defined as having log(SSFR)$\geq$-11[yr$^{-1}$] and $D_n4000<1.6$, in each of the four environments versus the mean Voronoi cell density of the galaxies in those environments. The colours indicate separate bins in stellar mass, with dwarf galaxies in blue and massive galaxies in red. The horizontal error bars, which are excluded from the dwarf galaxy points for clarity, indicate the standard deviation of a Gaussian fit to the Voronoi cell density distribution in each environment. The vertical error bars are 1$\sigma$ errors from a bootstrapping method described in Section \ref{sfr_vs_environ}.}
\label{fig:sfrduo}
\end{figure*}

Figure \ref{fig:sfrduo} (Left) plots SFR versus stellar mass for galaxies in the cluster, group, filament, and void environments, overlaid with lines of constant SSFR. As we would expect from the established SFR-density relation at low-$z$, the denser environments are host to a significant fraction of quiescent galaxies, which are found below the log(SSFR)=-11 [yr$^{-1}$] line in the four panels of Figure \ref{fig:sfrduo} (Left). Conversely, we find a large fraction of the galaxies at low-density environments are SF. As stated in Section \ref{galex}, 5 per cent of our sample of supercluster members do not lie in areas mapped by {\it GALEX}, and we therefore lack sensitivity to total SFRs for these galaxies. The galaxies not mapped by {\it GALEX} are excluded from Figure \ref{fig:sfrduo}. To examine the differences between star-formation activity in each environment quantitatively, and separate from dependence on mass, we calculate the fraction of galaxies that are SF in each environment separately for dwarf and massive galaxies. Figure \ref{fig:sfrduo} (Right) plots the fraction of SF galaxies as a function of the mean density (from the Voronoi cell densities) of each of the four environments. We include horizontal error bars by fitting a Gaussian to the cell density distribution in each environment, and defining the 1$\sigma$ error as the standard deviation of the best-fitting Gaussian to each. To get vertical error bars in our plot of SF fractions, we measured bootstrapped errors by resampling the galaxy populations of each environment 1000 times, allowing for repeats, and we estimate the 1$\sigma$ error on the SF fraction in each environment by measuring the standard deviation of the 1000 SF fractions from the resampled sets.

We find a steady decline in the fraction of SF galaxies as a function of density from the void to the cluster environment, with 96\% (65\%) of the dwarf (massive) galaxies SF in the voids and 25\% (11\%) of the dwarf (massive) galaxies SF in the clusters. The `mass quenching' effects can be seen separately from `environment quenching', as the mass-dependent effects are reflected in the different normalisations of the trends in Figure \ref{fig:sfrduo}. This result strongly indicates that environmental factors play a role in quenching star formation activity in higher-density environments, and that the group environment is undoubtedly a part of this quenching. Interestingly, we also see a statistically significant decline, in both dwarf and massive populations, in the fraction of SF galaxies between the void and the filament, which is not commonly considered as a site of environmentally-driven galaxy evolution. Indeed, most studies of the evolution of galaxies versus environment focus on over-dense regions, and would group everything we classify as filament and void galaxies together as `the field'. 

There are some ways in which our technique for identifying LSS could introduce an artificial trend showing a decline in the fraction of SF galaxies in the filament relative to the void. If our selection of $l_{crit1}$, which defines the threshold between cluster/group and filament galaxies, is too short, then there will be cluster or group galaxies mistakenly identified as being part of the filament environment. And if cluster and group environments are intrinsically composed of a higher fraction of quiescent galaxies, then their mistaken inclusion in the filament population would introduce a slight bias towards lower SF fraction in the overall filament population. Furthermore, if the filament environment contains small, compact groups of galaxies, consisting of fewer than eight galaxies in close proximity to each other, these small groups would be ignored by the MST algorithm when identifying groups, and they would likely end up labeled as filament galaxies. If these small groups also contain a higher fraction of quiescent galaxies than the overall filament population, then they will contaminate the SF fraction of the filament population. However, in Appendix \ref{appendix_a} we show that our results, including the lower SF fraction in the filament relative to the void, are still seen over a wide range of $l_{crit}$ values that result in unnaturally large, and unnaturally small, cluster and group populations, and that our results hold when requiring a minimum of just four galaxies (rather than eight) per structure identified with the MST.

\subsubsection{SFR Distribution versus Environment}\label{ssfr_distrib_vs_environ}

One method of examining possible environmental impact on SF activity is to compare the SFR or SSFR distributions of SF galaxies in each environment. Doing so can help reveal the timescale of any environmentally-driven quenching mechanism(s), as observing a distinct change in SSFR distribution of SF galaxies versus environment would indicate that quenching can occur on relatively long timescales. \citet{wetzel2012} examined the SSFR distributions of satellite galaxies (where `satellite' refers to a galaxy that is not the central galaxy of a host halo) of stellar masses $M_*\geq$10$^{9.7}\msun$ residing in haloes of masses $M_{halo}=10^{11.5}-10^{15}\msun$, selected using a `group-finder' algorithm with SDSS data, to determine whether there are trends in the quenching of SF activity of satellites related to the host halo mass, satellite galaxy mass, or the halo-centric radius. \citet{wetzel2012} show bimodal SSFR distributions with progressively lower fractions of SF galaxies residing in haloes of increasing mass, for all satellite mass bins in their sample, and they find no evidence of a change in the SSFR distribution of SF galaxies across their sample. Similarly, \citet{peng2010} found no evidence that the SSFR distribution of SF galaxies depends on environment, using a sample of galaxies taken from the SDSS over the redshift range 0.02$<z<$0.085. However, the \citet{peng2010} SDSS sample is biased towards massive galaxies, as at $z=0.085$ it is complete only to $M_*\sim$10$^{10.4}\msun$. They apply a weighting scheme to lower-mass galaxies by 1/V$_{max}$ to correct for incompleteness, but this correction is only valid if the dwarf galaxies they detect at all redshifts are an unbiased sampling of the dwarf galaxy population.

One of the most dangerous potential pitfalls when analysing galaxy activity versus environment can come from a dependence of galaxy mass on environment. As Figure \ref{fig:stmass_distribution} shows, the cluster and group environments have higher relative fractions of massive galaxies compared to dwarf galaxies. The well-established trend of galaxy `downsizing', whereby the more massive galaxies formed, and also had their SF activity quenched, at earlier epochs compared to dwarf galaxies \citep{cowie1996} can complicate the interpretation of galaxy SF activity versus environment, because denser environments also tend to be traced by more massive galaxies \citep{kauffmann2004}. Therefore, it is possible to mistakenly identify the effects of downsizing, and the well-documented SFR-$M_*$ correlation \citep{noeske2007, elbaz2007}, for an environmentally-driven effect when examining populations of galaxies with different mass distributions. We avoid this pitfall in our analysis by examining trends of SF activity separately for dwarf and massive galaxies in all our environments, and we can therefore ensure that our results are not merely tracing the effects of downsizing and secular evolution of galaxies.

\begin{figure*}
\centering
	\includegraphics[width=5.5in]{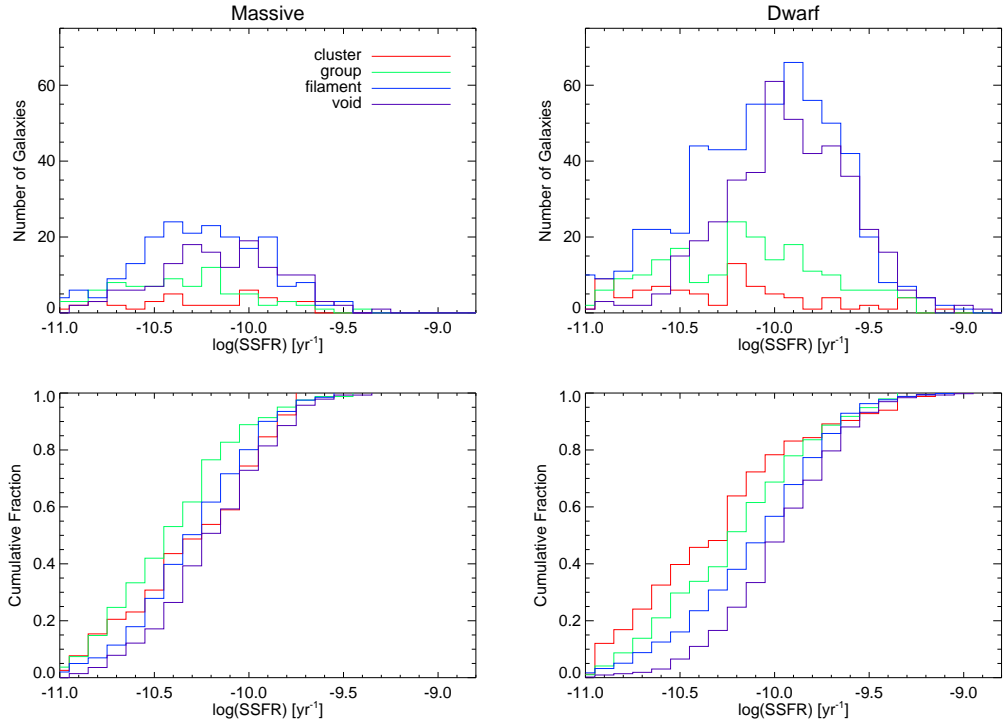}
	\caption{(Top) The differential distribution of SSFRs for SF galaxies in the four environments of the Coma Supercluster. (Bottom) The cumulative distribution of SSFRs for SF galaxies. The two left panels give the distributions for massive galaxies, and the two right panels are for dwarf galaxies.}
	\label{fig:ssfr_distrib}
\end{figure*}

Our sample, which is complete to $M_*\geq$10$^{8.5}\msun$, allows us to probe the SSFR distribution of SF galaxies to lower stellar mass regimes than in the previous studies. In Figure \ref{fig:ssfr_distrib} we show that the differential and cumulative distributions of SSFRs for SF galaxies in each of the four environments of the Coma Supercluster, separated by dwarf and massive galaxies. The SSFR distributions for SF dwarf galaxies in lower-density environments appear to be peaked at higher SSFRs than the dwarf SF galaxies of higher-density environments.

To quantitatively compare our SSFR distributions, we apply a Mann--Whitney U test \citep{mannwhitney1947}, which computes the probability $P_U$ (on a one-sided scale of 0--0.5) that two sets of points are drawn from an identical distribution, to the SSFRs for all pairs of environments in the supercluster. This means that if $P_U<0.0014$, we can reject the null hypothesis that two sets of data points are drawn from the same distribution with at least a 3$\sigma$ significance. When we contrast each pair of SSFR distributions for SF galaxies in the four environments, we find that only the SSFRs of the cluster and group environments are consistent with being drawn from the same distribution. See Table \ref{tab:sfr_distrib_pu} for the $P_{U_{SSFR}}$ values. However, differences in the SSFR distributions of SF galaxies in these four environments could arise due to an environmental dependence on SFR distribution of SF galaxies or due to differences in the underlying stellar mass distributions. To attempt to reconcile these two interpretations, we have also calculated the U statistic between the SFRs of SF galaxies in the four environments, and the U statistics comparing the SFRs of dwarf and massive SF galaxies in each environment. These additional sets of U statistics are also presented in Table \ref{tab:sfr_distrib_pu}. 

The SFR distributions of all SF galaxies in the different environments, with U statistics denoted by P$_{U_{SFR}}$ in Table \ref{tab:sfr_distrib_pu}, are much more statistically likely to have been drawn from the same distribution when compared to the SSFRs, which indicates that there is a definite dependence on mass that is affecting these results. When we compare the distributions of SFRs for SF galaxies separated into dwarf and massive bins, denoted P$_{U_{SFRd}}$ and P$_{U_{SFRm}}$, respectively, we find that the statistical differences in these SFR distributions are being driven mainly by the dwarf population. We cannot rule out the null hypothesis that all the SFR distributions for massive SF galaxies in all environments are being drawn from the same distribution, whereas for dwarf SF galaxies only the cluster-group distributions are consistent with being drawn from the same distribution. 

This result could be indicative of a number of things. Perhaps the environments of groups and clusters feature physical conditions which can drive gradual quenching of dwarf SF galaxies. This gradual quenching would have to occur on long enough timescales that we would be capable of seeing a statistically distinct `green valley' SF population at lower SFRs than the dwarf SF galaxies in lower-density environments. However, the process(es) acting to slowly quench dwarf galaxies in groups and clusters does not seem to be affecting the massive SF galaxies, at least not to a sufficient degree that the SFR distribution of massive SF galaxies in these higher-density environments is statistically distinct from the distribution seen at lower-densities. It is difficult to draw strong conclusions from this result, as the sample sizes of SF dwarf and massive galaxies in the high-density environments are small enough that we may be seeing artifacts of small number statistics. A future study, expanded greatly in overall sample size but still sensitive to low-mass galaxies, may be needed to better address the environmental dependence we find in our SFR distributions. Nevertheless, our results for massive SF galaxies agree with previous studies \citep[e.g. ][]{peng2010, wetzel2012} that have examined the environmental dependence of SFR distributions of massive SF galaxies. This result also underscores the importance of having greater galaxy mass completeness in surveys aimed at examining trends in galaxy evolution versus environment, as being limited to only massive galaxies will lead to very different conclusions about the environmental dependence of SFRs of SF galaxies.

\begin{table}
 \centering
  \caption{Mann--Whitney U test probabilities comparing the SSFR and SFR distributions for SF galaxies (with log(SSFR)$\ge$-11[yr$^{-1}$], $D_n4000<1.6$, and positions within \textit{GALEX} coverage areas) in the four environments. $P_{U_{SSFR}}$ compares the SSFR distributions of all SF galaxies in the Coma Supercluster environments. $P_{U_{SFR}}$ compares the SFR distributions of SF galaxies in all environments. $P_{U_{SFRd}}$ and $P_{U_{SFRm}}$ compare the SFR distributions for dwarf and massive SF galaxies, respectively, in all environments. Numbers in bold indicate cases where we cannot rule out the null hypothesis of the SSFRs or SFRs being drawn from the same distribution with at least a 3$\sigma$ significance.} \label{sfr_distrib_pu}
  \begin{tabular}{@{}lllll@{}}
  \hline
   Environments  &  $P_{U_{SSFR}}$  &  $P_{U_{SFR}}$ & $P_{U_{SFRd}}$ & $P_{U_{SFRm}}$ \\
 \hline
 Cluster-Group     &  \textbf{0.17}        &  \textbf{0.37}      & \textbf{0.026}     & \textbf{0.046} \\
 Cluster-Filament  &  $<$0.0014   &  \textbf{0.042}     & $<$0.0014 & \textbf{0.13} \\
 Cluster-Void      &  $<$0.0014   &  $<$0.0014 & $<$0.0014 & \textbf{0.47} \\
 Group-Filament    &  $<$0.0014   &  \textbf{0.0085}    & $<$0.0014 & \textbf{0.11} \\
 Group-Void        &  $<$0.0014   &  $<$0.0014 & $<$0.0014 & \textbf{0.0036} \\
 Filament-Void     &  $<$0.0014   &  \textbf{0.0046}    & $<$0.0014    & \textbf{0.014} \\
\hline
\label{tab:sfr_distrib_pu}
\end{tabular}
\end{table}

\subsection{Colour versus Environment}\label{colour_vs_environ}

\begin{figure*}
\centering
\begin{tabular}{cc}
\includegraphics[width=3.4in]{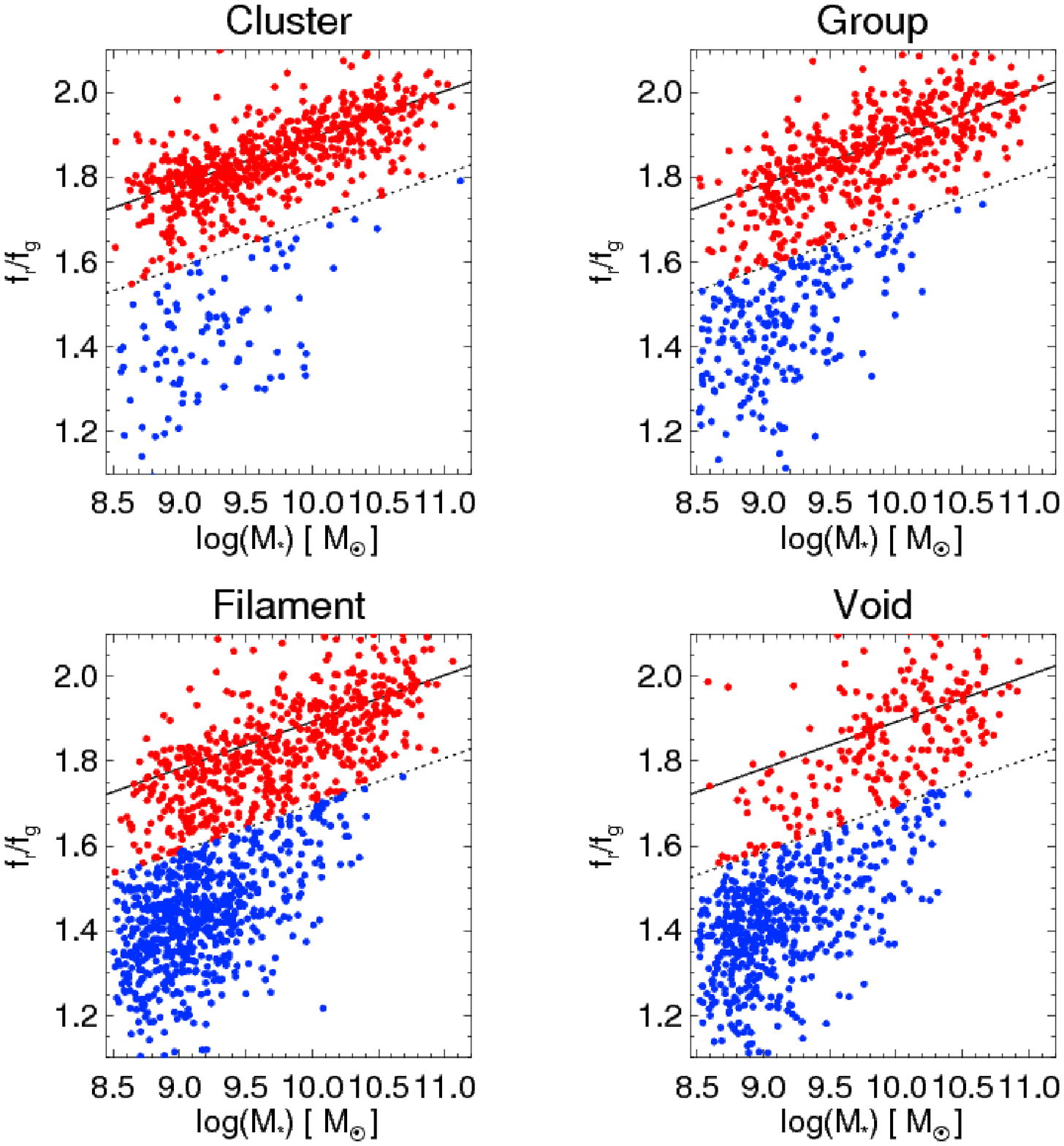} 
\includegraphics[width=3.4in]{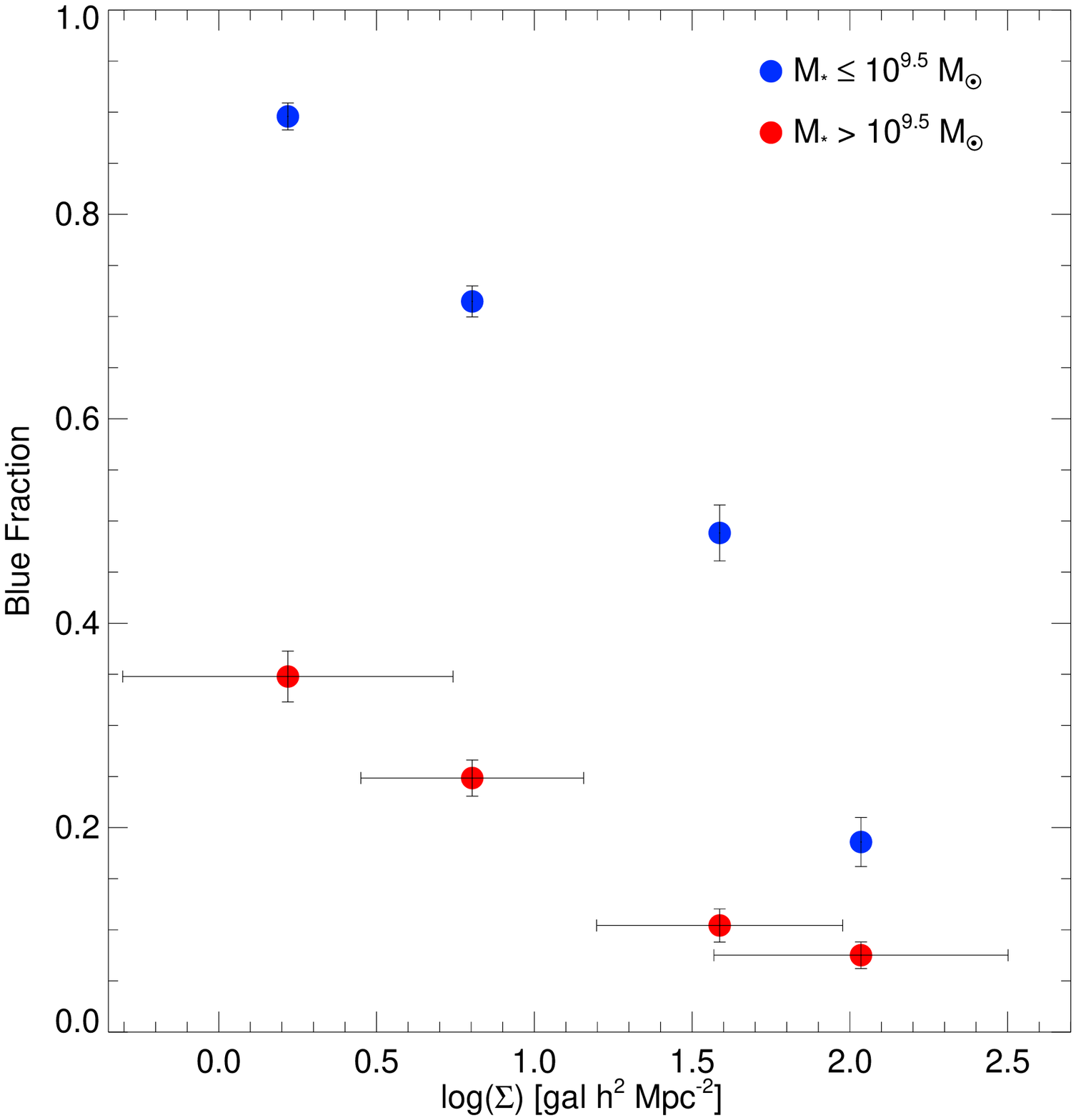}
\end{tabular}
\caption{(Left) Flux ratios of $r$ to $g$ versus stellar mass for galaxies in each of the four environments described in section \ref{mst}. The solid line denotes a fit to the red sequence, and the dashed line is the 2$\sigma$ threshold below the red sequence, where we separate `red' from `blue' galaxies (for more details, see Section \ref{colour_vs_environ}). (Right) Fraction of blue galaxies, defined using the $f_{r}$/$f_{g}$ versus stellar mass criterion defined in the left figure, in each of the four environments (void, filament, group, and cluster, in order of increasing mean density). The colours indicate separate bins in stellar mass, with dwarf galaxies in blue and massive galaxies in red. The horizontal error bars, which are excluded from the dwarf galaxy points for clarity, indicate the standard deviation of a Gaussian fit to the Voronoi cell density distribution in each environment. The vertical error bars are 1$\sigma$ errors from a bootstrapping method described in Section \ref{sfr_vs_environ}.}
\label{fig:colourduo}
\end{figure*}

To examine the dependence of galaxy colour on environment, we consider $g-r$, or the equivalent flux ratio $f_{r}/f_{g}$. Plotting the ratio of $f_{r}/f_{g}$ versus stellar mass for the entire sample reveals the familiar bimodal galaxy distribution \citep[e.g., ][]{baldry2004}, with a distinct red sequence and a blue cloud. To clearly separate red and blue galaxies, we start by fitting a line to the distribution of only the galaxies whose spectra are dominated by an older stellar population, by using the $D_n4000$ index. After fitting a line to the colour versus stellar mass distribution of galaxies with $D_n4000\ge1.6$, we define the standard deviation of the red sequence population about that best-fitting line as the 1$\sigma$ scatter. Then we define blue galaxies as those which are lower than the 2$\sigma$ threshold below the red sequence line. Figure \ref{fig:colourduo} (Left) shows our colour versus stellar mass plots for each of the four environments in the Coma Supercluster, and demonstrates our selection of red and blue galaxies.

\begin{figure}
	\includegraphics[width=3.2in]{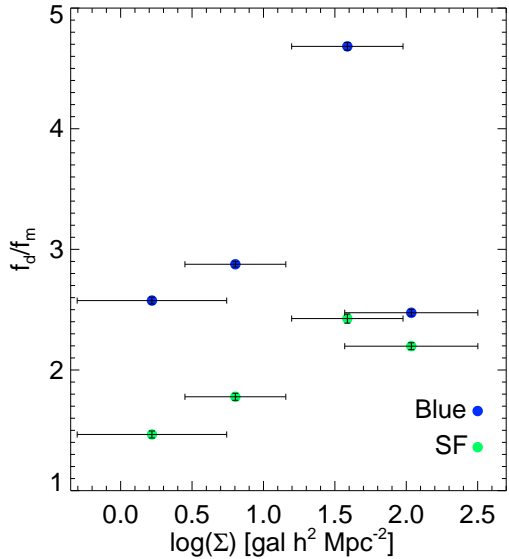}
	\caption{Ratios of dwarf-to-massive fractions of SF (green) and blue (blue) galaxies versus environment. In all environments the dwarf population is bluer and has a higher fraction of SF galaxies than their massive counterparts, but the differences between the mass bins is far more pronounced in the blue fractions than in the SF fractions.}
	\label{fig:dwarf_to_m_ratios}
\end{figure}

When calculating the blue fraction of galaxies in each environment of the Coma Supercluster, we take a similar approach to our SF fractions and examine trends with respect to galaxy mass by computing the blue fraction of galaxies in each environment separately for dwarf and massive galaxies. Figure \ref{fig:colourduo} (Right) presents the fraction of blue galaxies in each environment and mass bin as a function of the mean Voronoi cell density in the four environments. The horizontal and vertical error bars are calculated in the same manner as in Figure \ref{fig:sfrduo} (Right). We find a trend of steadily-declining blue fractions at increasing densities, for both mass bins, but with systematically lower fractions, especially for massive galaxies, when compared to the SF fractions of Figure \ref{fig:sfrduo}. For dwarf galaxies, the blue fraction tends to be lower than the SF fraction by less than 10\% in all environments, with the greatest deviation being in the group environment. The blue fraction for massive galaxies shows a much more dramatic decline compared to the SF fraction, with blue fractions about 50 per-cent lower in the lower-density environments. Figure \ref{fig:dwarf_to_m_ratios} more clearly demonstrates the differences between our fractions of SF and blue galaxies as a function of environment, by plotting the ratio of dwarf-to-massive fractions. 

\begin{figure*}
	\includegraphics[width=4.7in]{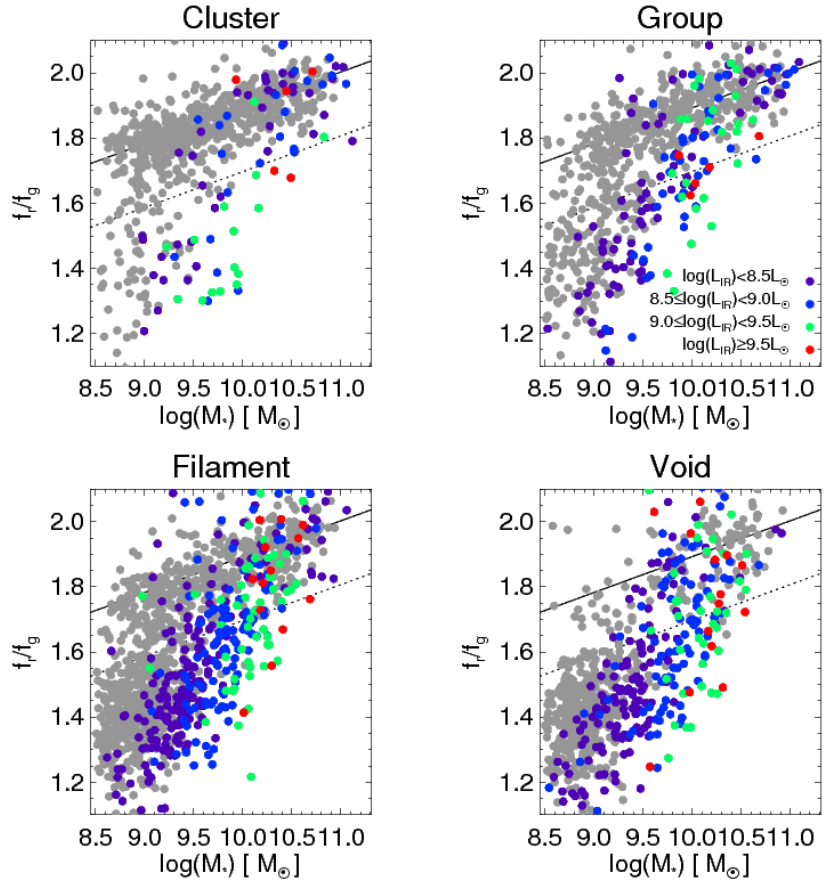}
	\caption{Flux ratios of r to g versus stellar mass for galaxies in each of the four environments, like in Figure \ref{fig:colourduo}, but with galaxies colour-coded by their IR luminosities inferred from WISE [22]. The grey points indicate non-detections in WISE [22], while the purple, blue, green, and red colours indicate progressively greater L$_{IR}$. Note that we are excluding the L$_{IR}$ information for any galaxy dominated by an AGN, as derived in \citet{brinchmann2004}.}
	\label{fig:color_withirlum}
\end{figure*}

The differences between our fractions of SF galaxies and blue galaxies most likely arise as a consequence of internal extinction within the galaxies in our sample. Galaxies with high dust content may experience significant reddening of their optical colours, and the magnitude of their extinction is proportional to both the inclination angle and mass of the galaxy, as demonstrated in Appendix A of \citet{gavazzi2013}. Since our study includes both un-obscured and obscured measures of SFR, the SF activity of these galaxies is `corrected' for the loss due to extinction. However, the optical colours are not corrected for this effect in our study (only for foreground Milky Way extinction), which can potentially introduce a significant bias to one's interpretation of the optical colour-density relation. Figure \ref{fig:color_withirlum} demonstrates the tendency for dusty, SF galaxies, particularly at high mass, to have redder optical colours, by showing that many massive dusty galaxies end up on the red sequence of a $g-r$ colour versus $M_*$ distribution.

Our blue fraction results, if considered alone, would lead one to conclude that the massive galaxies of the Coma Supercluster exhibit a dramatically lower incidence of recent SF activity, compared to the dwarf galaxies, in all environments, and that the overall environmentally-driven quenching of massive galaxies is a very weak effect. Our results do show, as seen in Figure \ref{fig:dwarf_to_m_ratios}, that a higher fraction of dwarf galaxies are actively SF and bluer than massive galaxies in all environments. However, the degree to which SF activity is suppressed as a function of galaxy mass, based on the optically-derived blue fractions alone, presents a misleading picture that is not in agreement with the results of our SF fractions.

\subsection{Post-starburst Galaxies versus Environment}\label{kpa_vs_environ}

Post-starburst, or k+A, galaxies show spectral signatures of having undergone significant SF activity in the recent past ($<$1--1.5Gyr ago), but have no substantial ongoing star-formation \citep{dresslergunn1983, couchsharples1987, poggianti1999}. For a recent review, see \citet{poggianti2009a}, and references therein. k+A galaxies are commonly identified as having strong Balmer absorption features but little-to-no emission lines in their spectrum. These galaxies can serve as a valuable signpost to identify the regions where significant SF activity was recently abruptly quenched, and therefore they provide a useful comparison to our analysis, which so far has focused primarily on contrasting populations of currently star-forming to currently quiescent galaxies.

We identify k+A galaxies following the definition of \citet{mahajan2010}, by selecting galaxies with EW(H$\delta)>$3\AA\ (indicating strong absorption) and EW(H$\alpha)>$-2\AA\ (meaning weak-to-no emission). Note that the signs indicating emission or absorption, positive EW for absorption and negative EW for emission, are reversed from what was used in \citet{mahajan2010}, reflecting a switch in the convention used for SDSS spectral line measurements. For EW(H$\alpha$) measurements we use the ratio of the H$\alpha$ flux to the H$\alpha$ continuum obtained for SDSS DR10 in \textbf{GalSpecLine}. To obtain H$\delta$ equivalent widths we used the Lick index measurement H$\delta_A$, originally proposed by \citet{wortheyottaviani1997} and available in SDSS DR10 from \textbf{GalSpecIndx} \citep{kauffmann2003}. Note that the H$\alpha$ equivalent widths, which are explicitly available in \textbf{GalSpecLine} in addition to the flux and continuum measurements for H$\alpha$, are less accurate than taking the ratio of flux-to-continuum because the equivalent widths are obtained without simultaneous fitting of all lines, and so H$\alpha$ is blended with [NII] unless one uses the flux-to-continuum ratio (C. Tremonti, private comm.). 

For our sample of 3505 supercluster galaxies, we find 62 having the spectral signatures of k+A galaxies. The k+A population in our sample is mostly dwarf galaxies predominantly concentrated in the high-density environments, with 32 and 9 k+A galaxies in the cluster and group environments, respectively. Figure \ref{fig:all_qu} shows the distribution of k+A galaxies, and demonstrates visually that these galaxies tend to favour high-densities. Table \ref{tab:kpa} presents the general environmental distribution and average masses of k+A galaxies in the Coma Supercluster. Our sample differs considerably from that of \citet{mahajan2010}, who presented a catalogue of 110 k+A galaxies in the Coma Supercluster using the same selection criteria. We have determined that the inconsistencies between our post-starburst catalogues stems from the significant differences in the spectral line measurements between those utilised by \citet{mahajan2010} and those which we are using. We elaborate on this comparison in Appendix \ref{appendix_c}.

\begin{table}
 \centering
  \caption{Statistics of k+A galaxies in the Coma Supercluster. Column 2 gives the number of k+A galaxies. Column 3 shows the fraction of galaxies in each environment that are k+As. Column 4 gives the log of the mean stellar mass of all galaxies in these environments, while Column 5 gives the log of the mean stellar mass of all k+A galaxies in the environment.} \label{kpa}
  \begin{tabular}{@{}lllll@{}}
  \hline
   Environ.  &  N$_{k+A}$  &  f$_{k+A}$ & log($<$M$_*$$>$) & log($<$M$_*$$>_{k+A}$)\\
             &             &            &     [$\msun$]     &  [$\msun$] \\
 \hline
 Cluster  &  32  &  0.043   &   10.03 & 9.13 \\
 Group    &  9   &  0.013   &   10.06 & 9.03 \\
 Filament &  19  &  0.015   &   9.86  & 9.13 \\
 Void     &  2   &  0.0026  &   9.81  & 9.62 \\
 All      &  62  &  0.018   &   9.94  & 9.15 \\
\hline
\label{tab:kpa}
\end{tabular}
\end{table}

Our results are in agreement with other studies of k+A galaxies \citep[e.g.,][]{poggianti2004} which have found that these post-starburst galaxies are rare at low-$z$, and those that are seen in the local universe tend to be dwarf galaxies. In contrast, clusters at $z=0.5$ were found by \citet{poggianti2004} to host a significantly larger fraction of k+A galaxies, and the higher-$z$ clusters had much more substantial populations of massive post-starburst galaxies. These differences between low- and high-$z$ likely reflect the global decline in SF activity and the effect of downsizing, whereby in the local universe massive starbursting galaxies are exceedingly rare.

\section{Discussion}\label{discussion}

\subsection{What's Driving Pre-Processing in the Coma Supercluster?}

Based on our results, in agreement with numerous other studies, galaxy groups at low-$z$ are prominent sites of galaxy transformation, as roughly half of all galaxies in groups in the Coma Supercluster are quiescent and red. The transformation mechanism(s) in the group environment can be a combination of different factors, most notably galaxy harassment and ram-pressure stripping, depending on the group host halo mass, dynamics of the group, and the possible presence of a hot intra-group medium (IGM). A detailed study of the velocity dispersions of group members and possible extended X-ray detections would help us to better assess which physical mechanism is dominating the transformation of the group galaxies throughout the Coma Supercluster, or whether the dominant physical mechanism is a function of group mass or even galaxy mass, but such a study is beyond the scope of this work.

\begin{figure*}
\centering
\begin{tabular}{c}
\includegraphics[width=6.5in]{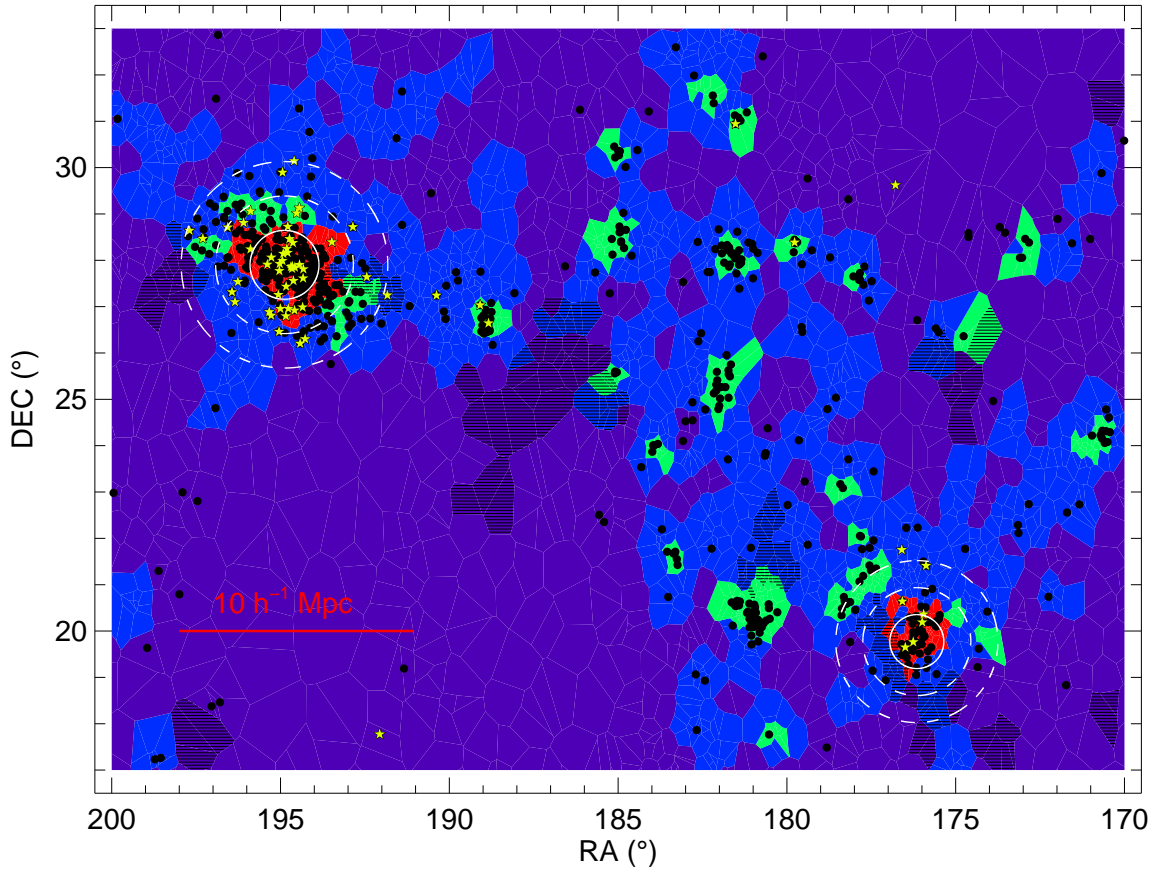}
\end{tabular}
\caption{Plot of the MST-defined galaxy environments, as in Figure \ref{fig:mst_tiers}, but with the positions of quiescent galaxies indicated by black dots. The yellow stars indicate the positions of k+A galaxies. The white solid circles indicate the virial radii of the two clusters, and the concentric dashed circles indicate the 2R$_{vir}$ and 3R$_{vir}$ radii. The cells with hatch marks indicate the galaxies not in \textit{GALEX} coverage areas, which have been excluded from the analysis of SF activity in this work.}
\label{fig:all_qu}
\end{figure*}

We have also found a marked decline in the fraction of SF and blue galaxies between the void and filament populations, and for dwarf and massive galaxies alike. This decline leads us to consider what process(es) may be responsible for quenching galaxies outside of the cores of clusters and groups, and some insight can be gained by examining the spatial distribution of quiescent galaxies in the supercluster. In Figure \ref{fig:all_qu} we plot the positions of the all quiescent and k+A galaxies in the Coma Supercluster. The quiescent galaxies in the filament tend to be clustered very near to the outskirts of galaxy clusters and groups, but at projected separations that put them well beyond the virial radii of the closest massive halo. 

There are several possible reasons we find a pronounced build-up of quiescent, and k+A, galaxies in the filament environment at projected separations of $\sim$2--5 times the virial radii of massive clusters and groups, and we can turn to recent work with simulations to help identify plausible scenarios. \citet{bahe2013}, hereafter B2013, recently presented some results from the GIMIC project \citep{crain2009}, which carried out higher-resolution hydrodynamical simulations on portions of the Millennium Simulation \citep{springel2005} that span a wide dynamic range of environments. B2013 focused on one particular result of the GIMIC simulations: that environmentally-affected galaxies are `observed' as far as $\sim$5 times the virial radius of the nearest massive cluster or group halo. The simulations presented in B2013 are particularly apt for comparing to this work, because the range of cluster/group halo masses being examined, with $M_{tot}\sim$10$^{13}$--10$^{15}\msun$, closely matches the mass distribution of structures we identify in the Coma Supercluster. The simulation shows that for a massive cluster, like A1656, at $z\sim$0 we should expect that as many as two-thirds of the galaxies within the virial radius have been pre-processed prior to entering the cluster, and perhaps 25--33 per-cent of the galaxies observed out to 5$R_{vir}$ have been pre-processed. For the most part, the pre-processing of these galaxies has come from gravitational interactions in group-mass haloes prior to their accretion onto the cluster, although there is some evidence that ram pressure stripping can occur due to galaxy interactions with gas in the group environment and in infalling regions of the extreme cluster outskirts (B2013).

Another possible reason for some of the quiescent galaxies being observed at large projected radii from massive haloes is overshooting of cluster/group members (B2013). Galaxies which are on elliptical orbits about massive haloes, and may have already had a pericentric passage within the virial radius of the host halo, could be observed as far as $\sim$2--3R$_{vir}$ in projected separation from the halo centre. These overshooting galaxies need not be pre-processed prior to their initial infall, as passing through the cluster core at least once can lead to significant ram pressure stripping and tidal effects imposed on the galaxy. According to the simulations of B2013, roughly half the galaxies observed at R$\sim$2R$_{vir}$ could be overshooting galaxies on highly elliptical orbits, but that fraction drops steeply and becomes negligible by R$\sim$3R$_{vir}$. 

\subsection{Comparison with Other Studies}

\subsubsection{Coma Studies}
Since the Coma Supercluster has been the target of numerous prior studies examining the dependence of environment on galaxy evolution, we have taken care to examine how our techniques for mapping the environment, and the conclusions we draw from our techniques, compare to previous work in this field. We apply the environment mapping method presented in \citet{gavazzi2010}, hereafter referred to as G2010, which uses a fixed-aperture-based technique to measure the volume density field over the Coma Supercluster, to our data set to compare our techniques side-by-side. In Appendix \ref{appendix_b} we describe our application of the density mapping technique of G2010 in greater detail, and show that our VT-based surface densities correlate strongly with the volume densities obtained by the methods of G2010.

We also demonstrate in Appendix \ref{appendix_b} that our segregation of galaxies into cluster, group, filament, and void populations in general agrees with the delineations made in G2010 based on local under- or over-density. However, in Figure \ref{fig:density_comparison_duo} we find a non-negligible fraction of galaxies in our MST-defined environments that are distributed amongst the other surrounding `environment bins' as defined by G2010. Some galaxies that we defined as cluster members end up re-distributed into the `cluster outskirts/group galaxy' category of G2010, and some of our group galaxies are shuffled into the `filament' bin of G2010, which is to be expected since these components of the cosmic web lack clearly-defined boundary demarcations. However, it is worthwhile to note that the differences in environmental grouping of our study and G2010 are not entirely attributed to the use of different density thresholds, nor are the differences completely due to our use of 2D densities versus 3D. Our assignment of a galaxy to a given environment is not based purely upon the local density surrounding that galaxy, but also on whether the galaxy is continuously connected to a nearby structure by sufficiently short MST branches. For example, a galaxy located on a cluster outskirts might be classified in a lower-density bin by the G2010 criteria, but if it is connected to the rest of the cluster by projected branches of length $l\leq l_{crit1}$, then we would classify it as a cluster galaxy.

Table \ref{tab:environs_gavazzi} presents the basic statistics of the galaxy environments obtained when using the methods of G2010, which helps demonstrate how their environmental selection differs from our own. The primary difference between the G2010 environments and the present work (as given in Table \ref{tab:environs}) is that the population of cluster galaxies, and the area covered by those cluster galaxies, is much smaller by the G2010 criteria. The sharp drop in the average SSFR of the cluster population, as defined by G2010 criteria, relative to what we show in Table \ref{tab:environs} reflects the fact that the G2010 cluster galaxies reside only in the highest-density cluster core, where SFR is most strongly suppressed. The fact that the number of cluster galaxies is lower by 30\%, while the total surface area of the cluster galaxies drops by more than half is indicative that the galaxies that we identify on the cluster outskirts, which are at lower projected densities, are preferentially getting re-distributed into the `Group' and `Filament' populations in the G2010 designations. However, a large number of our group galaxies are also preferentially getting distributed into the `Filament' category of G2010.

\begin{table}
 \centering
  \caption{Environments of the Coma Supercluster defined using the technique of \citet{gavazzi2010}. Column 3 gives the total area of the Voronoi cells of the galaxies in each environment. Column 4 gives the mean Voronoi cell surface density of the galaxies in each environment. Column 5 gives the log of the average specific SFR (SSFR=SFR/M$_*$) of galaxies in each environment, taken as the sum of the SFRs of the galaxies in each environment (excluding galaxies which are dominated by an AGN, see Section \ref{sfrs}) divided by the sum of the stellar mass of all galaxies in that environment.} \label{environs_gavazzi}
  \begin{tabular}{@{}llllr@{}}
  \hline
   Environ.  &  N$_{gal}$  &  Area  &   $<$$\Sigma$$>$  &  log$(<$SSFR$>)$  \\
                &             &   (h$^{-2}$ Mpc$^{2}$) &  (h$^{2}$ Mpc$^{-2}$) &    [yr$^{-1}$]  \\
 \hline
 `Cluster' & 518  & 8.4  & 139.0 & -11.47  \\
 `Group'   & 845  & 59.3  & 36.7  & -11.17  \\
 `Filament'& 1373 & 436.3 & 9.5   & -10.63  \\
 `Void'    & 769  & 887.6 & 1.8   & -10.41  \\
\hline
\label{tab:environs_gavazzi}
\end{tabular}
\end{table}

Furthermore, we want to be sure that our results, in terms of the quenching of SF activity versus environment, remain largely unchanged when using the methods of G2010 to map the environment of the Coma Supercluster. Using the environment designations given in Table \ref{tab:environs_gavazzi}, when we examine the fraction of SF galaxies (dwarf and massive) we find the same overall trends of decreasing SF fraction at higher-density environments that we report in Figure \ref{fig:sfrduo}. The results of G2010 also showed an increasing fraction of early-type galaxies in higher-density environments, but they conclude that the bulk of the environmental-dependence is driven by the dwarf galaxies in their sample, and that massive galaxies have little dependence on their environment. However, because the G2010 study used $g-i$ colors, and was therefore prone to optical extinction bias for the colors of massive galaxies, it's not surprising that a much weaker environmental trend was seen for massive galaxies in G2010.

\subsubsection{Other Studies of Environmentally-Driven Effects}

\citet{scoville2013} examined the SF activity of galaxies as a function of local density in the two square degree COSMOS \citep{scoville2007} field over a wide redshift range (0.15$\leq$$z$$\leq$3.0) and split into 127 redshift bins using photometric redshifts from \citet{ilbert2013}. They calculate local galaxy density in each of the redshift bins using VT, as well as an adaptive smoothing technique for comparison, and show that early types with lower SFRs are more common at higher densities for redshifts below $z\sim$1.3. Furthermore, the degree to which the high-density population of galaxies differs from those at lower densities (in terms of the fraction of early-type galaxies, and in mean SFRs) is more extreme at lower redshifts. The \citet{scoville2013} study provides valuable insight into the evolution of the relation between galaxy properties and local density over cosmic time, but their sample features a smaller dynamic range of galaxy masses and densities. They are limited to galaxies that are more massive ($M_*\geq$10$^9\msun$), and to lower projected densities ($\Sigma\leq$100 galaxies Mpc$^{-2}$) than our current study.

The Local Cluster Substructure Survey \citep[LoCuSS;][]{smith2010} is examining 30 X-ray selected galaxy clusters at $z\sim$0.2, with the goal (among others) of spotting S0-progenitor spirals in the act of undergoing morphological and SFR changes on the outskirts of these clusters. By focusing on a large sample of clusters at one low-redshift bin, they are able to statistically sample the cluster-to-cluster variations in galaxy properties down to moderate-mass galaxies, and without being adversely affected by intrinsic changes in galaxies, and cluster evolution, as a function of redshift. The LoCuSS survey utilises Herschel PACS \citep{poglitsch2010} photometry extending to cluster-centric radii of $\sim$1.5 times the virial radius, and with sufficient sensitivity to detect bolometric IR luminosities L$_{IR}\ge3\times$10$^{10}\lsun$ for the $z\sim$0.2 clusters. Our sensitivity to L$_{IR}$ is almost two orders of magnitude deeper in the Coma Supercluster, and we are also able to detect signs of galaxy transformation extending out to several times the virial radius of clusters and groups in our survey.

The Galaxy And Mass Assembly \citep[GAMA;][]{baldry2010, driver2011} survey is built upon 300,000 galaxy spectra, obtained with the AAOmega multi-object spectrograph on the Anglo-Australian Telescope, for galaxies with r$<$19.8 mag over 290 deg$^2$ spread over three fields. In addition to their new spectroscopy, they have \textit{GALEX} UV photometry, UKIRT Infrared Deep Sky Survey \citep[UKIDSS;][]{hewett2006, lawrence2007} NIR imaging, SDSS spectra and optical photometry, and \textit{Herschel} Astrophysical Terahertz Large Area Survey \citep[H-ATLAS;][]{eales2010, rigby2011} FIR imaging over their survey region. When the completed survey is released, it will make an excellent complimentary data set to the current study, and will be capable of extending an analysis of galaxy properties versus environment, for similar mass range as our present sample, to $z\sim0.2$. A recent study by \citep{alpaslan2013} identifies filamentary structures in the GAMA fields out to $z\sim0.2$ using the MST on galaxy group positions, rather than on galaxy positions. They also use the MST on galaxy positions to identify diffuse `tendrils' traced by galaxies, and show that the structures they identify are also found in mock catalogues.

\section{Conclusions}\label{conclusions}

We have mapped the environments and calculated the projected density field of the Coma Supercluster in detail, identifying the regions with clusters, groups, filaments, and voids, using the two complementary techniques of Voronoi Tessellation and the Minimal Spanning Tree. The 3505 supercluster members with stellar masses M$_*\ge10^{8.5}\msun$ are thus split into 741 cluster, 716 group, 1292 filament, and 756 void galaxies, allowing us to study the properties of galaxies in these discrete environments. 

We measure SFRs across the entire supercluster down to 0.02$\sfr$ using WISE W4 and {\it GALEX} NUV bands, while taking care to avoid biasing our SFRs by excluding the IR contribution for galaxies dominated by an AGN and the NUV contribution for galaxies with spectral characteristics of an old stellar population. 

The fraction of galaxies that are actively star-forming (with log(SSFR)$\ge$-11[yr$^{-1}$]) is progressively lower in environments of increasing density. This trend holds for massive and dwarf galaxies, and is even found when comparing the void to the filament environment. The fraction of galaxies with blue $g-r$ colours, bluer than the red sequence population, declines steadily in environments with increasing density, in a manner which is similar to our trends in SF fraction versus environment. However, as our $g-r$ colours are susceptible to internal extinction, we find a striking decline in the blue fraction of massive galaxies when compared to the SF fraction. This result underscores the importance of utilising unbiased measures of galaxy properties whenever possible, or of applying necessary corrections to biased measures \citep[e.g., see the optical extinction corrections described in Appendix A of][]{gavazzi2013}.

We compare the SFR distributions of SF galaxies in all four environments, and find statistically distinct SFR distributions for dwarf galaxies for most environments, but for massive galaxies the SFR distributions in all environments are consistent with being drawn from the same distribution. This result suggests that the process(es) most effective at quenching massive galaxies may do so on shorter timescales than the process(es) which are most responsible for quenching dwarf galaxies, but we also may be prone to small number statistics. Nonetheless, our results underscore the importance of using a large baseline of galaxy masses when examining trends in galaxy evolution.

We identify 62 k+A, or post-starburst, galaxies in our Coma Supercluster sample based on SDSS spectroscopy, and find them to be predominantly dwarf galaxies in higher-density environments, which is consistent with similar studies of post-starburst galaxies at low-$z$. These k+A galaxies are primarily located near the core of the massive cluster A1656, which supports the hypothesis that these post-starburst systems have been recently quenched by ram pressure stripping or by tidal effects due to recent group-cluster merging.

The spatial distribution of quiescent galaxies in the Coma Supercluster confirms that the evolution of galaxies via pre-processing is extremely prevalent in the local universe. On average, galaxies in groups are about half as likely to be actively star-forming, when compared to galaxies in the void. Furthermore, we find a significant over-abundance of quiescent galaxies in filaments and groups on the outskirts of the massive cluster A1656 (at projected cluster-centric radii of $R\sim2-3 R_{vir}$). Simulations \citep[e.g.,][]{bahe2013} suggest that some of these galaxies are overshooting cluster members on elliptical orbits, but the majority of these quiescent galaxies on the cluster outskirts require a pre-processing scenario (with transformations that began in their host groups prior to infall, or from tidal interactions extending far beyond the cluster virial radius) to explain their presence. Our results agree with simulations of hierarchical clustering of DM haloes \citep[e.g.,][]{mcgee2009, delucia2012}, which suggest that most galaxies accreting onto massive clusters at $z\sim0$ have been affected by pre-processing prior to their arrival in the cluster environment.

\section*{Acknowledgments}
We thank our anonymous referee, who suggested numerous revisions and clarifications that improved the paper. This work benefitted greatly from discussions with Stacey Alberts, Matthew Ashby, Daniela Calzetti, Daniel Eisenstein, and Christy Tremonti.

This research has made use of the NASA/IPAC Extragalactic Database (NED) which is operated by the Jet Propulsion Laboratory, California Institute of Technology, under contract with the National Aeronautics and Space Administration. This research has also made use of NASA's Astrophysics Data System (ADS), and the NASA/IPAC Infrared Science Archive, which is operated by the Jet Propulsion Laboratory, California Institute of Technology, under contract with the National Aeronautics and Space Administration. This publication makes use of data products from the Wide-field Infrared Survey Explorer, which is a joint project of the University of California, Los Angeles, and the Jet Propulsion Laboratory/California Institute of Technology, funded by the National Aeronautics and Space Administration. Some of the data presented in this paper were obtained from the Mikulski Archive for Space Telescopes (MAST). STScI is operated by the Association of Universities for Research in Astronomy, Inc., under NASA contract NAS5-26555. Support for MAST for non-HST data is provided by the NASA Office of Space Science via grant NNX09AF08G and by other grants and contracts. Funding for the SDSS and SDSS-II has been provided by the Alfred P. Sloan Foundation, the Participating Institutions, the National Science Foundation, the U.S. Department of Energy, the National Aeronautics and Space Administration, the Japanese Monbukagakusho, the Max Planck Society, and the Higher Education Funding Council for England. The SDSS Web Site is http://www.sdss.org/. This research made use of Montage, funded by the National Aeronautics and Space Administration's Earth Science Technology Office, Computational Technnologies Project, under Cooperative Agreement Number NCC5-626 between NASA and the California Institute of Technology. The code is maintained by the NASA/IPAC Infrared Science Archive.

R. Cybulski was supported by the Smithsonian Astrophysical Observatory (SAO) Predoctoral Fellowship, and funded under NASA Contract 1391817 and NASA ADAP grant 13-ADAP13-0155. M.~S. Yun acknowledges support from the NASA ADAP grant NNX10AD64G. R.~A. Gutermuth gratefully acknowledges funding support from NASA ADAP grants NNX11AD14G and NNX13AF08G and Caltech/JPL awards 1373081, 1424329, and 1440160 in support of \textit{Spitzer} Space Telescope observing programs.

\appendix

\section{Robustness of Environment Mapping Technique}\label{appendix_a}

Here we explore to what degree our results are affected by large variations in the application of our techniques for mapping the density field and identifying LSS. A significant concern, especially given the degree to which we find quiescent galaxies in the filament clustering near the outskirts of groups and clusters (as seen in Figure \ref{fig:all_qu}), is that a small change in the choice for $l_{crit}$, which would alter the spatial extent of the galaxy cluster, group, and filament populations in our map, might significantly change our results. For example, if we use a larger value for $l_{crit1}$, and therefore have more galaxies included in our cluster and group populations, many of the filament galaxies which lie on cluster and group outskirts would end up absorbed into the clusters and groups themselves, which would raise the SF fraction in the filament population.

To be sure that our decline in the fractions of blue and SF galaxies with each progressively higher-density environment is not simply an artifact of our specific choices for $l_{crit1}$ and $l_{crit2}$, we ran our entire analysis over a very wide range of parameters. In Section \ref{mst} we used surface density thresholds of 40 and 10 galaxies h$^2$ Mpc$^{-2}$ to select the characteristic $l_{crit}$ lengths corresponding to the threshold between cluster/group and filaments, and filaments and voids, respectively. To test the robustness of our results to the particular choices of density thresholds, we vary the density threshold between cluster/group and filament galaxies between 100--20 galaxies h$^2$ Mpc$^{-2}$, in intervals of 20 galaxies h$^2$ Mpc$^{-2}$, and the threshold for filament and void galaxies between 12--4 galaxies h$^2$ Mpc$^{-2}$, in intervals of 4 galaxies h$^2$ Mpc$^{-2}$. Note that the extreme ends of the density thresholds being tested represent obviously unrealistic structures, i.e. with 100 galaxies h$^2$ Mpc$^{-2}$ only the inner-most core members of the clusters and groups are included in the cluster and group populations, while at 20 galaxies h$^2$ Mpc$^{-2}$ the cluster and group extent is unnaturally large. 

Another potential source of bias in our results comes from our requirement that at least eight galaxies be present in a continuously-connected substructure in order to be recognized by the MST algorithm. Recall that we chose eight galaxies as a minimum to reduce the occurrence of spurious groups of galaxies arising due to projection effects. However, we are therefore ignoring any galaxy group which has fewer than eight members, meaning that they would end up classified as either filament or void galaxies depending on the local galaxy density around them. So if small galaxy groups tend to be scattered throughout the filamentary regions of the Coma Supercluster, and less often in the void-like regions, and the galaxies within these small groups are more likely to be quiescent than similar galaxies in isolation, then we could be artificially decreasing the SF fraction of the filament relative to the void due to these small groups. To address this concern, we also tested the full range of densities described previously, but with a minimum of four galaxies per structure identified with the MST algorithm. This choice undoubtedly results in a large number of spurious galaxy `groups', but it should significantly mitigate any bias we are introducing into the filament population by ignoring small groups of galaxies.

\begin{figure}
	\includegraphics[width=3.0in]{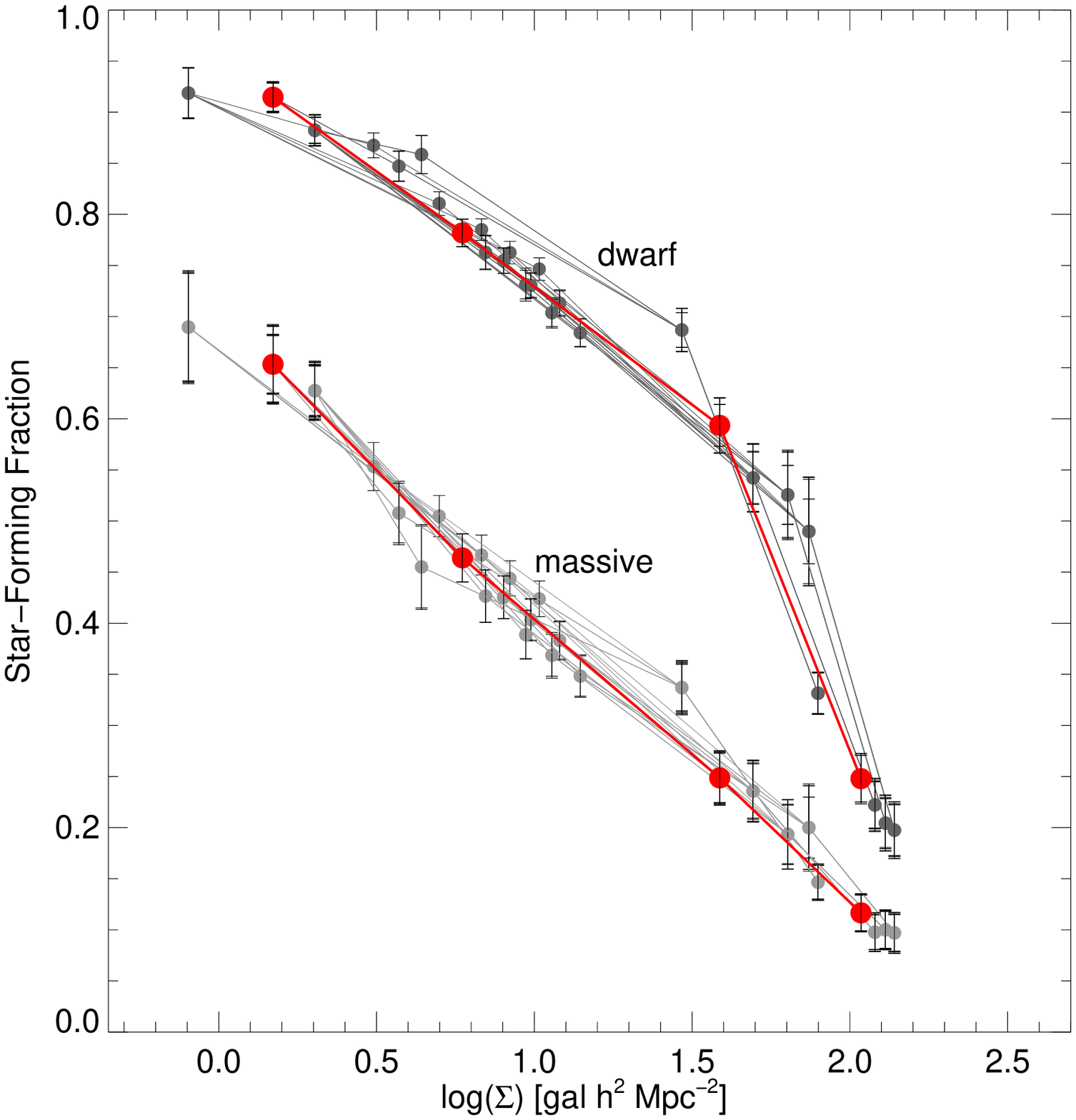}
	\caption{Fractions of SF galaxies versus the mean density of the four environments of the Coma Supercluster calculated using a large range in surface densities to define $l_{crit1}$ and $l_{crit2}$. The upper set of points, in dark grey, are the SF fractions of dwarf galaxies. The lower set of points, in lighter grey, are for massive galaxies. The red points indicate the fractions calculated using $l_{crit1}$ and $l_{crit2}$ reported in Section \ref{mst}.}
	\label{fig:big_sffract_set}
\end{figure}

We ran our MST code to separate galaxies into the four environment categories for all combinations of density thresholds described above, and with a minimum of eight and four galaxies per structure identified by the MST, and calculate the fraction of SF galaxies in each environment as a function of mean VT cell density for each trial run. Figure \ref{fig:big_sffract_set} presents the SF fractions as a function of environment for this range of tests, and as before (see Figure \ref{fig:sfrduo}) we have separated the samples into dwarf and massive galaxies. Despite the fact that our range of tests have significant differences in the distribution of galaxies into the four environments, and therefore they vary in the range of densities in each environment, we still find the same overall trends of declining SF fraction at progressively higher-density environments.

\section{Detailed Comparison with Previous Mapping of Coma Supercluster}\label{appendix_b}

Here we present a more thorough comparison of how our techniques for mapping the supercluster environment, and characterising local density, compare to previously-published studies for this field. Specifically, we compare with the work of G2010, who used a fixed-aperture technique to measure the volume density field in the Coma Supercluster. G2010 measure the local density around every galaxy using a cylinder of radius 1 h$^{-1}$ Mpc, and with a distance along the line-of-sight equal to $\pm$1000 km s$^{-1}$. 

There are a few minor differences that are worth noting before we apply their technique. Our survey volumes are very similar, but G2010 used a slightly larger range of line-of-sight velocities (4000$<cz<$9500 km s$^{-1}$ versus our 4495$\leq$$cz$$\leq$8973 km s$^{-1}$) and a marginally smaller angular area (420 sq degrees to our 480 sq degrees), and they supplement their SDSS-selected sample with 177 additional galaxies with spectroscopic redshifts in the literature. We are using only our SDSS sample, and we also exclude a small fraction of our sample at the lowest-masses to ensure that our sample is complete and unbiased over our range of stellar masses (see Section \ref{sdss}). Nevertheless, our galaxy samples agree, by number, to within 15\% and the mean volume density of galaxies, given our number of galaxies and total volumes, are nearly identical; we find 0.06 galaxies h$^3$ Mpc$^{-3}$ for our sample, and G2010 measure 0.05 galaxies h$^3$ Mpc$^{-3}$. Additionally, G2010 deal with edge effects by assuming constant boundary conditions (by dividing the density of any cylinder which falls partially outside their survey by the fraction of the cylinder which is inside the survey volume), whereas we have obtained an additional `buffer' several degrees wide surrounding our field, comprised of galaxies chosen with exactly the same selection criteria from the SDSS as the main sample (see Section \ref{vt}), which allows us to bypass any biases due to edge effects.

There is an additional significant deviation introduced by G2010 that is worth mentioning. They rightly point out that the extreme dynamic range of densities in the Coma Supercluster makes using densities measured in a fixed volume at every galaxy position problematic. The high velocity dispersions of galaxies in the clusters necessitate using a `longer' cylinder to encompass the true local density of galaxies, but using a longer cylinder over the entire supercluster population would result in less sensitivity to local density in more rarefied regions. The compromise of G2010 was to shrink the velocity dispersion of members of the clusters A1656 and A1367, by assuming that their transverse sizes ($\sim$2 deg and $\sim$1 deg for A1656 and A1367, respectively) also reflect their sizes along the line-of-sight. In effect, this modification is similar to using a longer cylinder to measure the volume of galaxies in the clusters compared to the rest of the supercluster.

\begin{figure*}
\centering
\begin{tabular}{cc}
\includegraphics[width=3.45in]{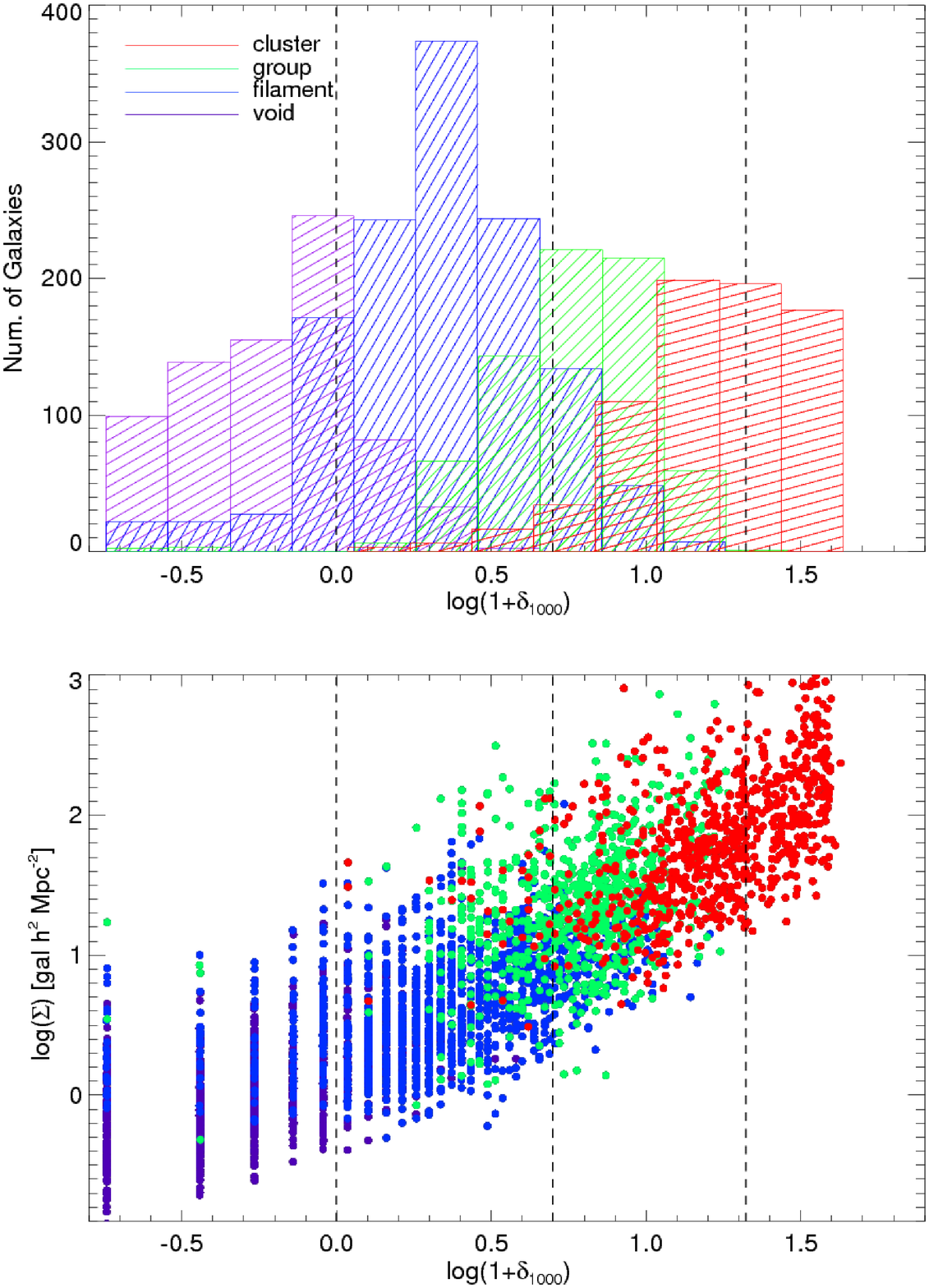} 
\includegraphics[width=3.45in]{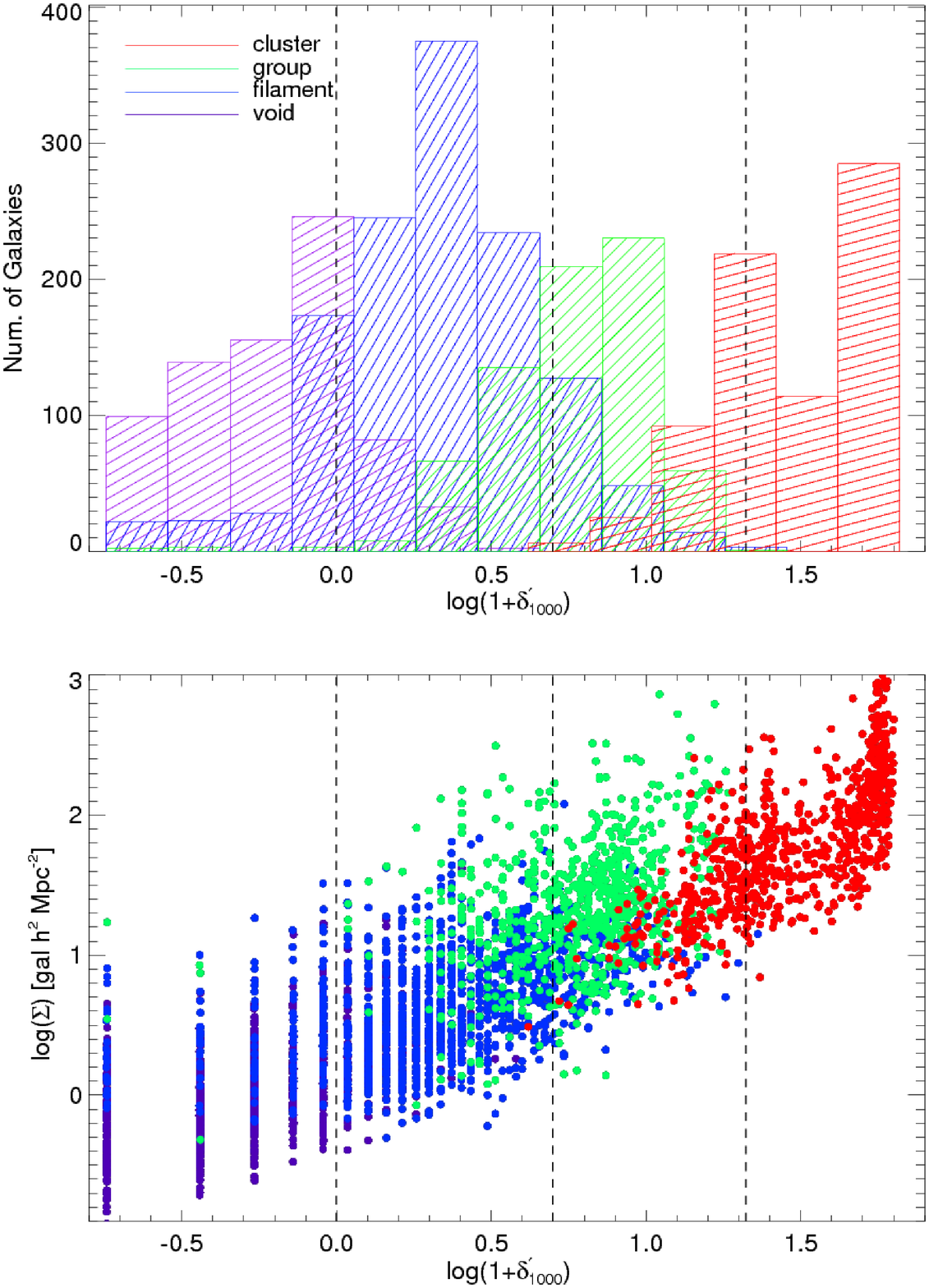}
\end{tabular}
\caption{(Left pair) Comparison of density distributions, and selection of environmental segregation, between our dual VT + MST approach and that of a `fixed aperture' cylinder of radius 1 h$^{-1}$ Mpc and half-length of 1000 km s$^{-1}$. The top left panel shows the differential distribution of galaxies which we classify (using the MST as described in Section \ref{mst}) as belonging to the cluster, group, filament, and void versus the volume density parameter from G2010, measuring the relative under- and over-density from a fixed aperture. The vertical dashed lines indicate the separation into environmental bins (roughly corresponding to void, filament, group, and cluster from left-to-right) used by G2010. The bottom left panel shows the correlation between our Voronoi-based surface densities to the volume densities calculated in the cylinder as described previously, with points colour-coded based on our MST-defined environment categories. The large gaps that appear on the left-hand side are due to the fact that the fixed aperture method counts discrete numbers of galaxies per unit volume. (Right pair) The same comparison, but in this case the fixed aperture densities were calculated exactly as in G2010, with the spread in line-of-sight velocities of members of the two clusters A1656 and A1367 reduced by assuming spherical symmetry in these two systems.}
\label{fig:density_comparison_duo}
\end{figure*}

For the purposes of this comparison with previous mapping techniques, we have measured the local volume density around every galaxy in our sample using a cylinder of fixed half-length equal to 1000 km s$^{-1}$ with two cases: 1) the velocities of all galaxies are exactly as reported in the SDSS and 2) velocities for cluster members are modified according to the prescription of G2010. Figure \ref{fig:density_comparison_duo} presents a comparison between our VT-based surface densities, and our MST-defined environmental segregation, with the volume densities calculated using these two approaches with fixed-aperture cylinders. Our definitions for galaxy environment largely agree, and they agree more so in the case in which the velocities of cluster galaxies are adjusted, but there is some overlap between different environments in adjoining classifications. When the velocity dispersion of the cluster galaxies is artificially lowered, we unsurprisingly see a large amplification in the local density for the cluster galaxies, and a much smaller fraction of cluster galaxies at intermediate densities.

\section{k+A Galaxy Follow-up}\label{appendix_c}

\begin{figure*}
	\includegraphics[width=4.5in]{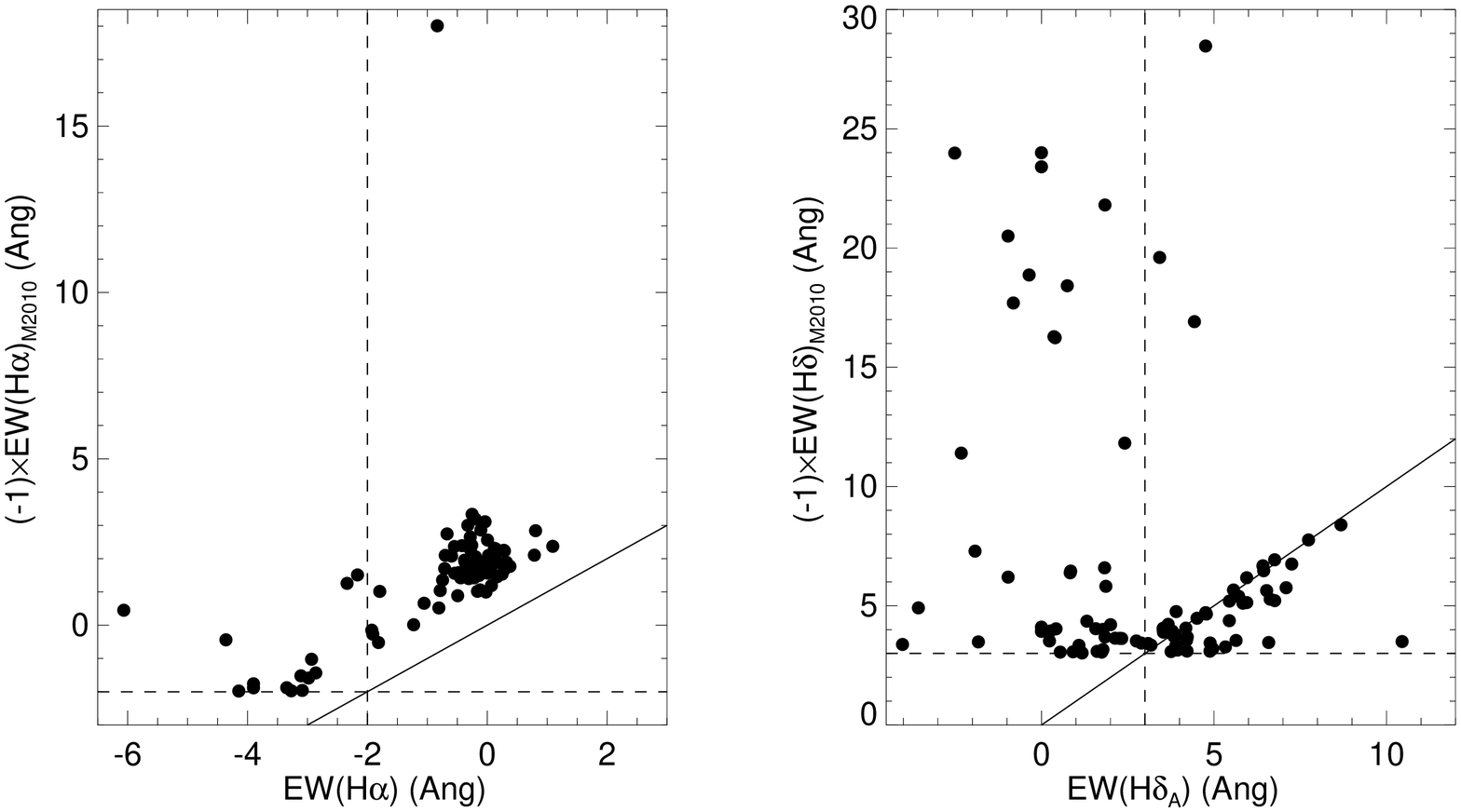}
	\caption{(Left) Comparison between the H$\alpha$ equivalent widths obtained for the k+A galaxies presented in M2010 versus the equivalent widths we obtain. The solid line is a 1:1 correlation, and the dashed lines indicate the -2 Angstrom threshold used as a minimum requirement in the selection of k+A galaxies. (Right) Comparison between the H$\delta$ equivalent widths of k+A galaxies in M2010 versus our equivalent widths. A factor of -1 is applied to the equivalent widths of M2010 to reflect the reversal in convention to denote absorption or emission with positive or negative numbers}.
	\label{fig:h_alpha_delta_compare}
\end{figure*}

\citet{mahajan2010}, hereafter M2010, presented a table of 110 dwarf k+A galaxies in the Coma Supercluster, which come from a parent supercluster galaxy sample very similar to that of the present work, and with an identical set of criteria to select the post-starburst galaxies (see Section \ref{kpa_vs_environ}). However, we find only 62 such k+A galaxies. We matched the published M2010 k+A catalogue to our own, and find that 91 of the 110 galaxies proposed by M2010 are indeed in our parent supercluster sample. However, only 41 of the 110 proposed k+A galaxies from M2010 show spectral characteristics, based on our SDSS line measurements, indicative of being k+A. Figure \ref{fig:h_alpha_delta_compare} shows the comparison between the M2010 EW(H$\alpha$) and EW(H$\delta$) measurements, and those from the present study, for the 91 proposed k+A galaxies matched to our sample. It's abundantly clear that there are significant differences in the spectral line measurements for both lines relevant for selecting k+A galaxies, with a prominent systematic offset for H$\alpha$ and a large scatter in the measurements of H$\delta$. 

The reasons for these inconsistencies stem from the different techniques used to obtain the line measurements. The data used in the M2010 catalogue come from the \textbf{SpecLine} products, but for the present work we use \textbf{GalSpecLine} and \textbf{GalSpecIndx}, which are based on the MPA-JHU analysis \citep[see][]{kauffmann2003}. For the \textbf{SpecLine} data products, the spectrum continuum is fit using a sliding mean/median filter (C. Tremonti, private comm.), while for \textbf{GalSpecLine} and \textbf{GalSpecIndx} the continuum is fit with stellar population synthesis models \citep{kauffmann2003, tremonti2004}. The differences in continuum fitting techniques are responsible for most of the systematic offset between H$\alpha$ measurements in Figure \ref{fig:h_alpha_delta_compare}, as the underlying H$\alpha$ from the stellar population has not been subtracted from the measurements used in M2010. There is also a factor of $(1+z)$ difference because the \textbf{SpecLine} measurements were not converted to the rest-frame, but for the redshift of Coma this is a relatively small effect. The differences seen in H$\delta$ come from a combination of the lack of robust continuum fitting and the fact that the \textbf{SpecLine} products measure H$\delta$ using a simple Gaussian. The H$\delta$ measurement used for our present study comes from the Lick index H$\delta_A$, first proposed by \citet{wortheyottaviani1997}, and also described in \citet{prochaska2007}, which is designed to optimally capture the H$\delta$ absorption feature in the atmospheres of A-type stars.

\footnotesize{

}

\label{lastpage}

\end{document}